\documentclass[12pt,letter]{article}

\usepackage{graphicx, epsfig, color, amssymb}
\def\eslt{E_T^{\rm miss}}
\def\to{\rightarrow}

\def\bi{\begin{itemize}}
\def\ei{\end{itemize}}

\def\sps1ap{SPS1a$^\prime$}
\def\c1p{C1$^\prime$}

\def\tb{\tilde b}

\def\tst{\tilde t}
\def\tsb{\tilde b}

\def\tg{\tilde g}

\def\tq{\tilde q}

\def\tw{\widetilde W}
\def\tz{\widetilde Z}
\def\alt{\lesssim}
\def\agt{\gtrsim}
\def\be{\begin{equation}}  
\def\ee{\end{equation}}  
\def\bea{\begin{eqnarray}}  
\def\eea{\end{eqnarray}}  
\def\beas{\begin{eqnarray*}}  
\def\eeas{\end{eqnarray*}}

\newcommand{\hepph}[1]{hep-ph/#1}

\newcommand{\lfig}{12 cm}

\begin{document}
\begin{titlepage}
\begin{flushright}
UH-511-1281-17
\end{flushright}

\vspace{0.3cm}
\begin{center}
{\large \bf
The Reach of the High-Energy LHC for Gluinos and Top Squarks \\
in SUSY Models with Light Higgsinos 
}\\ 
\vspace{1.cm} \renewcommand{\thefootnote}{\fnsymbol{footnote}}
{\large Howard Baer$^1$\footnote[1]{Email: baer@ou.edu }, 
Vernon Barger$^2$\footnote[2]{Email: barger@pheno.wisc.edu },
James S. Gainer$^3$\footnote[3]{Email: jgainer@hawaii.edu },
Hasan Serce$^1$\footnote[4]{Email: serce@ou.edu },
and Xerxes Tata$^3$\footnote[5]{Email: tata@phys.hawaii.edu }
}\\
\vspace{1.cm} \renewcommand{\thefootnote}{\arabic{footnote}}
{\it 
$^1$Dept. of Physics and Astronomy,
University of Oklahoma, Norman, OK 73019, USA \\
}
{\it 
$^2$Dept. of Physics,
University of Wisconsin, Madison, WI 53706, USA \\
}
{\it 
$^3$Dept. of Physics and Astronomy,
University of Hawaii, Honolulu, HI 96822, USA \\
}
\end{center}

\vspace{0.4cm}
\begin{abstract}
We examine the top squark (stop) and gluino reach of the proposed 33 TeV energy upgrade 
of the Large Hadron Collider (LHC33) in the Minimal Supersymmetric Standard Model (MSSM) 
with light higgsinos and relatively heavy electroweak gauginos.  
In our analysis, we assume that stops decay to higgsinos via $\tst_1 \to
t \tz_1$, $\tst_1 \to t\tz_2$ and $\tst_1 \to b\tw_1$ with branching
fractions in the ratio 1:1:2 (expected if the decay occurs dominantly
via the superpotential Yukawa coupling) while gluinos decay via
$\tg\to t\tst_1$ or via three-body decays to third generation quarks plus higgsinos. 
These decay patterns are motivated by models of natural supersymmetry where higgsinos 
are expected to be close in mass to $m_Z$, but gluinos may be as heavy as 
$5 - 6$ TeV and stops may have masses up to $\sim 3 $ TeV.
We devise cuts to optimize the signals from stop and gluino
pair production at LHC33. We find that experiments at LHC33 should be
able to discover stops with $> 5\sigma$ significance if $m_{\tst_1}
< 2.3 \ (2.8) \ [3.2]$~TeV for an integrated luminosity of 0.3 (1)
[3]~ab$^{-1}$. 
The corresponding reach for gluinos extends to 5 (5.5) [6]~TeV. 
These results imply that experiments at LHC33 should be able to discover 
at least one of the stop or gluino pair signals even with an integrated luminosity of
0.3~ab$^{-1}$ for natural SUSY models with no worse than 3\% electroweak
fine-tuning, and quite likely both gluinos and stops for an integrated
luminosity of 3~ab$^{-1}$. 

\noindent 

\end{abstract}

\end{titlepage}

\section{Introduction}
\label{sec:intro}

In spite of the fact that no direct evidence for superpartners has as
yet emerged in Large Hadron Collider (LHC) data, weak scale
supersymmetry (SUSY) arguably remains the most promising extension of
the Standard Model (SM). The remarkable ultra-violet properties of
softly-broken SUSY theories tame the radiative corrections in the Higgs
sector and thus provide a resolution~\cite{ghp} of the big hierarchy
problem that emerges when the SM is embedded in a Grand Unified Theory
(GUT). The MSSM with weak scale
superpartners is indirectly supported by several observations: 1)~the
measured values of gauge couplings, evolved to high scales, appear to
unify within the MSSM but not in the SM~\cite{gauge}; 2) the top quark
is heavy enough to radiatively drive electroweak symmetry breaking~\cite{rewsb}; 
and 3)~the measured Higgs boson mass~\cite{lhc_h}, which could
have been as high as several hundred GeV in the SM, lies in the
relatively narrow window $m_h \lesssim 135$~GeV 
(which is support for the MSSM if the MSSM
remains valid to $Q\sim M_{\rm GUT}$~\cite{mhiggs}).

The search for SUSY has long been one of the important items on
the agenda of the ATLAS and CMS experiments at the LHC. The absence of a
signal in the data has been interpreted as lower limits on various
sparticle masses.  The colored superpartners, the squarks and gluinos,
are the most stringently constrained, but there are also limits on the masses of 
electroweakly-produced charginos and neutralinos. For the most
part, these limits are obtained in simplified models, usually assuming
direct decays to the lightest SUSY particle (LSP) which is
taken to be the neutralino. From an analysis of 36~fb$^{-1}$ of data at
the ($13$ TeV) LHC, the ATLAS and CMS collaborations have reported the following 
95\% CL bounds.
\bi
\item $m_{\tq} \agt 1600$~GeV from a search for squark pair production 
(assuming 8 degenerate species of light squarks, a massless LSP 
 and decoupled gluinos~\cite{lhc-strong}).
\item $m_{\tg}\agt 1850-2000$~GeV from a search for gluino pair production
(assuming $\tg \to q\bar{q}\tz_1$ with a massless LSP and 
decoupled squarks~\cite{lhc-strong}).
\item $m_{\tw_1}=m_{\tz_2} \agt 550-600$~GeV from a search for wino pair production
(assuming $\tw_1 \to \tz_1 W$ and $\tz_2 \to \tz_1 Z$~\cite{lhc-wino}).
\ei

In many models the third generation squarks, especially $\tst_1$, are
considerably lighter than other squarks. 
In this case, the gluino would dominantly decay via $\tg \to t\bar{t}\tz_1$ 
and perhaps also via $\tg \to b\bar{b}\tz_1$ or $\tg\to tb\tw_1^\pm$. 
Under this motivation, the LHC collaborations have also searched for signals from 
stop pair production and also for gluinos dominantly decaying to 
third generation quarks. 
Such events would be rich in $b$-tagged jets. 
The absence of a signal above backgrounds has led to the following 95\% CL bounds 
(again for nearly massless LSPs).
\begin{enumerate}
\item $m_{\tst_1} \agt 950-1050$~GeV (assuming $\tst_1 \to t\tz_1$ or $Wb\tz_1$-- 
there is a similar bound on the bottom squark)~\cite{lhc-stop}, and
\item $m_{\tg} \agt 1960-2050$~GeV (assuming $\tg \to t\bar{t}\tz_1$
~\cite{lhc-glthird}).
\end{enumerate}
Considering an LSP with finite mass only mildly degrades these bounds
unless the LSP is relatively close in mass to the parent sparticle.

As the LHC continues to accumulate more data, the ATLAS and CMS experiments
will probe even larger superpartner masses. 
With an integrated luminosity of 300~fb$^{-1}$ -- 
expected to be accumulated at $\sqrt{s}=14$~TeV by the end of LHC Run 3 -- 
experiments should be able to probe gluinos with a
$5\sigma$ significance out to $\sim 1.9$~TeV ($\tg\to q\bar{q}\tz_1$)
and stops to about 950~GeV~\cite{upgrade,gershtein}. 
The 95\% CL reach is typically 300-400 GeV higher.
At the ``High Luminosity LHC'' (HL-LHC) where an
integrated luminosity of 3000~fb$^{-1}$ is anticipated, these reaches
extend to about 2300~GeV (gluino) and 1200~GeV
(stop)~\cite{upgrade,gershtein,atlaswiki}.
In a recent study, we have shown that with stringent selection cuts, 
the $5\sigma$ gluino discovery potential of (the $14$ TeV) 
LHC extends to 2.4 (2.8)~TeV, for an integrated
luminosity of 300 (3000)~fb$^{-1}$ if the gluino decays via $\tg \to
t\tst_1 \to t\bar{t}\tz_1$~\cite{us-14}. 

While colored sparticle searches are usually expected to probe most
deeply into SUSY parameter space at hadron colliders, for integrated
luminosities larger than 300~fb$^{-1}$ electroweak-ino pair production
becomes competitive, or even dominant for the case that the gluino
decays democratically into all generations.  For instance, wino pair
production via $pp\to \tw_2(\to W^\pm\tz_{1,2})\tz_4(\to W^\pm
\tw_1^{\mp})X)$ followed by decay to same-sign dibosons ($W^\pm W^\pm$)
may probe more deeply into parameter space than gluino production in the
well-motivated class of {\it natural SUSY} models with light higgsinos
and gaugino mass unification~\cite{lhcltr,lhc}.

If SUSY is not discovered at the HL-LHC, then the search for SUSY may
require new facilities.  An electron-positron linear collider with
$\sqrt{s}> 2\,m_{\rm higgsino}$ is particularly well suited~\cite{ilc}
for SUSY discovery since the requirement of light higgsinos has been
argued to be a very robust feature of natural SUSY models~\cite{nsusy,ltr,rns}. 
In this paper, however, we examine the SUSY reach
of the proposed {\it energy upgrade} of the LHC: a $pp$ collider
operating at $\sqrt{s}=33$~TeV with an anticipated integrated luminosity
of $\sim 1$~ab$^{-1}$, referred to hereafter as
LHC33~\cite{helhc}.\footnote{We will also consider the scenario where
LHC33 obtains 3~ab$^{-1}$ of integrated luminosity.}  We examine in
detail the LHC33 discovery reach for gluinos and stop pair production,
focussing on the case where $\tg \to t\tst_1$ and $\tst_1 \to t+ X$ or
$\tst_1 \to bX$ each with a branching fraction of 50\% and where the
decay products of $X$ are essentially invisible.\footnote{In a previous
study~\cite{jamieplb} we had shown that the gluino reach of LHC33
extended beyond 5~TeV, the exact value depending on the stop mass. In
this paper we re-evaluate the reach while improving our cuts for gluino
pair events.  Our evaluation of the stop reach is new.}

We will defer a detailed discussion of our motivation for focussing
on gluino and stop searches (with the assumed decay patterns) until
Sec.~\ref{sec:implications}.  Briefly, we will see that in a wide
variety of natural SUSY models, the gluino and stop masses are {\em
bounded from above} by naturalness conditions and that these
superpartners decay as we have assumed.  These upper bounds -- that
$m_{\tg}\alt 5-6$ TeV and $m_{\tst_1}\alt 3$ TeV -- imply that these
sparticles may well lie beyond the reach of HL-LHC.  Moreover, in
well-motivated SUSY models such as mirage-mediation with a compressed
spectrum of gauginos, electroweak-ino signals may also be beyond the
reach of HL-LHC.  For a definitive test of natural SUSY at a hadron
collider, an energy upgrade to the vicinity of $\sqrt{s}\sim 30$ TeV
seems necessary to probe the natural parameter space.  We find
that experiments at LHC33 will definitively be able to discover at least
one of these superpartners over the entire parameter space of natural
SUSY models, and both over much of the allowed parameter space.  While
we find these conclusions highly compelling, the reader who does not
subscribe to our naturalness criteria should view this paper instead
simply as an analysis of the LHC33 reach for gluino and stop
pair production in models with decays to light higgsinos.

Before proceeding further we should mention how our analysis differs
from the one existing study~\cite{gl33} of the reach for gluinos at
LHC33 that we are aware of. To our knowledge there is no detailed
corresponding study for stops.\footnote{In Ref.~\cite{gershtein}
Fig.~1-24 shows that LHC33 should be able to probe stops via the
single lepton channel, assuming $\tst \to t\tz_1$. Unfortunately, no
details are given.} In Ref.~\cite{gl33} the authors examine the discovery
reach of gluinos for various scenarios including $\tg \to t \tst_1$ of
interest to us. They focus, however, on the same-sign dilepton event
channel that is possible if the two (of four) leptonically decaying tops
have the same sign of electric charge. The rate for the signal is then
suppressed by the square of the leptonic branching fraction for top
quark decays, and also by the fact that such events are not possible if
$\tst_1 \to b\tw_1$, and the daughters of the chargino are too soft to
be efficiently detectable. Our strategy is, instead, to allow the fully
inclusive sample of gluino events requiring just hard tagged $b$-jets
and large $\eslt$ in the events as we did in an earlier analysis for the
HL-LHC. We will see that this leads to a significantly larger reach than
in Ref.~\cite{gl33}.

The remainder of this paper is organized as follows.
Sec.~\ref{sec:simulation} explains the procedures used to simulate the
relevant signal and background processes and also lists the reactions
that we have simulated.  Sec.~\ref{sec:event_selection} describes the
analyses used, in particular explaining how we arrive at our choice of
cuts for both the stop and the gluino signal.  In
Sec.~\ref{sec:reach}, we use these analyses to make projections for the
discovery and exclusion reach, in terms of the stop and gluino masses,
at LHC33.  The implications of these projected discovery and exclusion
reaches for natural SUSY models are discussed in
Sec.~\ref{sec:implications}. Finally, we summarize our main results and
present our conclusions in Sec.~\ref{sec:conclusions}.

\section{Event Simulation}
\label{sec:simulation}
We begin by describing how we simulate gluino and stop signal
events at LHC33 as well as the various SM backgrounds listed in
Sec.~\ref{subsec:processes_simulated}.

\subsection{Event Generation}
\label{subsec:simulation_procedure}

LHC33 events were generated using {\sc MadGraph}~2.3.3~\cite{madgraph}
interfaced to {\sc PYTHIA} 6.4.14~\cite{pythia} via the default
MadGraph/PYTHIA interface with default parameters for showering and
hadronization.  Detector simulation is performed by {\sc Delphes} using the
default Delphes~3.3.0~\cite{delphes} ``CMS'' parameter card with the 
modifications listed below.

\begin{enumerate}

\item We set the electromagnetic calorimeter (ECAL) energy resolution to
  $3\%/\sqrt{E} \oplus 0.5\%$ and the hadronic calorimeter (HCAL) energy
  resolution to be $80\%/\sqrt{E} \oplus 3\%$ for $|\eta| < 2.6$ and
  $100\%/\sqrt{E} \oplus 5\%$ for $|\eta| > 2.6$, where ``$\oplus$'' denotes
  combination in quadrature. These are typical of the values used for
  LHC calorimeters and do not assume any significant improvement at LHC33.

\item We turn off the jet energy scale correction.

\item We utilize the anti-$k_T$ jet algorithm~\cite{Cacciari:2008gp}
with $R = 0.4$ rather than the default $R = 0.5$.  (Jet finding in
Delphes is implemented via {\sc FastJet}~\cite{Cacciari:2011ma}.)  We
consider only jets with transverse energy satisfying $E_T(jet) > 50$ GeV
and pseudorapidity satisfying $|\eta(jet)| < 3.0$ in our analysis.  The
choice of $R = 0.4$ in the jet algorithm is made, in part, to facilitate
comparison with CMS $b$-tagging efficiencies~\cite{CMS:2016kkf}
described below.

\item We have developed our own module for jet flavor association to
implement the ``ghost hadron'' procedure~\cite{Cacciari:2007fd} that
allows decayed hadrons to be assigned to jets in an unambiguous manner;
we use this module to determine whether jets contain $B$ hadrons.  If a
jet contains a $B$ hadron (in which the $b$ quark decays at the next
step of the decay) with $|\eta| < 3.0$ and $E_T > 15$ GeV, then we
identify this $b$-jet as a ``truth $b$-jet''.  We tag $b$-jets with
$|\eta| < 1.5$ with an efficiency of $60\%$ and assume that light quark
and gluon jets with $|\eta| < 1.5$ can be mistagged as $b$-jets with a
probability of $1/150$ for $E_T < 100$ GeV, $1/50$ for $E_T > 250$ GeV
and a linear interpolation for $100$ GeV $ < E_T < 250$
GeV.\footnote{These values are based on ATLAS studies of $b$-tagging
efficiencies and rejection factors in $t\bar{t}H$ and $WH$ production
processes~\cite{ATLASb}.}  We have checked~\cite{us-14} that our $b$-jet
tagging algorithm yields good agreement with the $b$-tagging
efficiencies and mistag rates in Ref.~\cite{CMS:2016kkf}.
While these $b$-tagging parameterizations were developed for the ($14$
TeV) LHC, we expect that they represent reasonable starting points for
LHC33 phenomenology.  More detailed studies, beyond the scope of this
paper, will ultimately be needed to understand 
$b$-tagging at LHC33.

\item We do not perform ``tau tagging'', {\it i.e.}, identifying objects
as taus, as our analyses does not involve the tagging of hadronic taus
and we do not want to ``lose'' $b$-jets where the $B$ hadron in the
jet decays to a tau lepton.
\end{enumerate}

The lepton identification criteria that we adopt are the default for
this version of Delphes, namely leptons with $E_T> 10$~GeV and within
$|\eta_\ell| < 2.5$  are
considered isolated if the sum of the transverse energy of all other
objects (tracks, calorimeter towers, etc.) within $\Delta R = 0.5$ of
the lepton candidate is less than $10\%$ of the lepton $E_T$.

\subsection{Processes Simulated}
\label{subsec:processes_simulated}

As mentioned in Sec.~\ref{sec:intro}, our
goal is to map out the LHC33 reach for gluinos and stops, assuming
that these are the only colored superpartners that are kinematically
accessible. For definiteness, we assume that gluinos decay via $\tg \to
t\tst_1$, and that the stop decays via $\tst_1 \to t \tz_{1,2}$ or $\tst_1
\to b\tw_1$ with branching ratios of 0.25, 0.25 and 0.50,
respectively. These branching ratios are typical for models with light
higgsinos and relatively heavy gauginos, where the decay dominantly
occurs via the top quark Yukawa coupling~\cite{lhc}. Unless the bino and
wino are fortuitously also rather light, the visible decay products of
the daughter $\tz_2$ and $\tw_1$ are generally very soft because the
higgsino mass gap is typically 5-20~GeV. In our analysis of the LHC33
reach, we will assume that the higgsinos are not detectable in the experimental
apparatus. We should thus regard our results as conservative since the
additional (especially leptonic) debris from the neutralino and chargino
decays could be potentially used to further ``beat down'' SM backgrounds. The
associated gluino top-squark production cross section is negligible,
essentially because the top quark density in the proton is very small.  

Gluino pair production at LHC33 thus leads to $tttt$ + $\eslt$,
$tttb$ + $\eslt$, $ttbb$ + $\eslt$ and $bbbb$ + $\eslt$ events, where the $\eslt$
arises the fact that the daughter higgsinos are essentially invisible.
Likewise, stop pair production leads to $tt+\eslt$, $tb+\eslt$ and
$bb+\eslt$ events. SUSY events are thus signalled by the production of
2-4 very hard $b$-jets (not all of which would be tagged as $b$-jets) and
very large $\eslt$.

The dominant SM backgrounds come from SM processes with top and bottom
quarks in the final state, with the physics $\eslt$ arising primarily
from neutrinos from the decays of top quarks and $Z$ bosons. With this
in mind, we simulated backgrounds from $t\bar{t}$, $t\bar{t}b\bar{b}$,
$b\bar{b}Z$, and $t\bar{t}t\bar{t}$ production and from single top
production.\footnote{In a previous study at the 14~TeV LHC~\cite{us-14},
we had shown that backgrounds from vector boson production ($V$+ jet
production and $VV$+jet production) were very efficiently removed after
requiring two hard $b$-jets and large $\eslt$.} In our evaluation of the
background from top pair production, we veto $t\bar{t}$ events with more
than two truth $b$-jets, to avoid double counting (as we also explicitly
simulate $t\bar{t}b\bar{b}$).  When simulating our $t\bar{t}$,
$t\bar{t}b\bar{b}$, single top, and $b\bar{b}Z$ backgrounds, we generate
events in bins of generator-level missing transverse energy, $\eslt$.
Using weighted events from this procedure yields a more
accurate simulation of the high tail of the $\eslt$ distribution for
these background processes, which is essential for determining the rates
from background processes after the very hard $\eslt$ cuts that we use
for the extraction of the signal: see Sec.~\ref{sec:event_selection}.

\begin{figure}[tbp]
\begin{center}
\includegraphics[width=\lfig,clip]{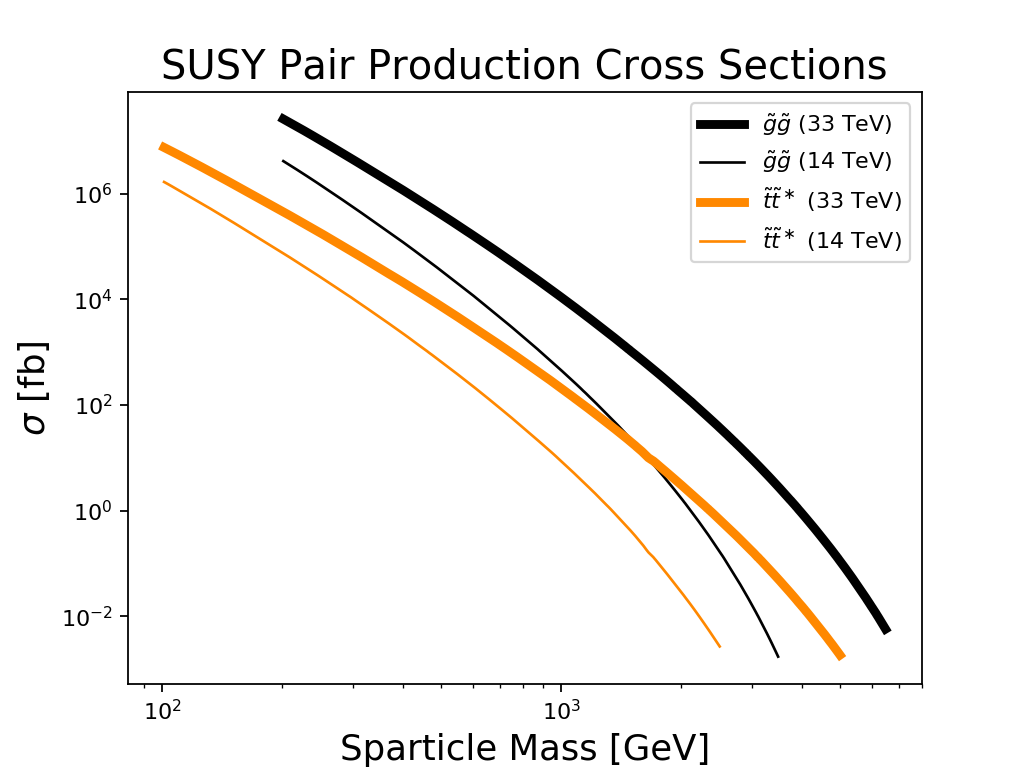}
\caption{ Total NLO+NLL cross sections for gluino (black) and stop
(lighter/ orange) production at LHC33 (thick curves) and, for comparison, at
the $14$ TeV LHC (thin curves).
\label{fig:total_xsec_33_14_TeV}}
\end{center}
\end{figure}

In our MadGraph simulation, we normalize the $2\to2$ pair production
cross section for our gluino pair production signals and our stop pair
production signals, for various stop and gluino masses, to the NLO + NLL
(next-to leading logarithm) values from the LHC SUSY cross section
working group~\cite{LHC_SUSY}.  We display these total cross sections,
as well as the corresponding cross sections at the ($14$ TeV) LHC in
Fig.~\ref{fig:total_xsec_33_14_TeV}. We see that $\agt 100$ gluino
(stop) pair events should be produced at LHC33 if $m_{\tg}$
($m_{\tst_1}$) $< 5 \ (3)$~TeV, assuming an integrated luminosity of
1~ab$^{-1}$.
The cross section for our top pair production background was normalized
to $5449.6$ pb, following ATLAS-CMS recommended
predictions\footnote{This is the cross section for a top mass of $173.2$
GeV, which we find by interpolation from the provided cross
sections.}~\cite{LHCTopWG_ttbar} for the LHC33, which are
ultimately based on Hathor v2.1~\cite{Hathor}.  As we were unable to find
NLO $K$-factors for the other background processes, we have used the same
$K$-factor values that we used for the $14$ TeV analysis in Ref.~\cite{us-14}.
These are $1.3$ for $t\bar{t}b\bar{b}$, following
Ref.~\cite{Bredenstein:2010rs}\footnote{ Larger $K$-factors for this
process are obtained when a dynamic scale choice is not
employed~\cite{bred2}.}; $1.5$ for $b\bar{b}Z$ production, following
Ref.~\cite{Cordero:2009kv}; $1.27$ for the $t\bar{t}t\bar{t}$ backgrounds,
following Ref.~\cite{Bevilacqua:2012em}, and $1.11$
for single-top cross sections, following
the ($14$ TeV) ATLAS-CMS recommended predictions~\cite{LHCTopWG} which
are ultimately based on the Hathor v2.1 program~\cite{Hathor}.

\section{Event Selection}
\label{sec:event_selection}
In this section, we develop sets of cuts to separate gluino pair and 
top-squark pair production events from SM backgrounds 
with the goal of optimizing the reach of LHC33. 
We first discuss the stop pair signal in Sec.~\ref{subsec:stop_analysis}
and then turn to the gluino pair production signal in Sec.~\ref{subsec:gluino_analysis}.

\subsection{Stop Analysis}
\label{subsec:stop_analysis}

In order to determine the discovery reach for stops at LHC33,
we must first develop a set of cuts to separate the stop
pair production signal events from various SM backgrounds. The reach will be
approximately optimized by tuning the cuts for the extraction of a
signal from a heavy stop. With this in mind and given the stop pair
production rate in Fig.~\ref{fig:total_xsec_33_14_TeV}, we adopt a
simplified model benchmark point with $m_{\tst_1}=2.75$~TeV, for which
$\sigma_{\tst_1\tst_1} \sim 300$~ab (corresponding to 100-300 events
before analysis cuts depending on the integrated luminosity)  
to enhance the stop signal over the background.
For definiteness, we adopt a higgsino mass parameter $\mu = 150$ GeV,
though we expect the results to be only weakly sensitive to this value (as
long as $|\mu|\ll m_{\tst_1}$).
The masses of relevant particles are given in Table~\ref{tab:stop_bm}.  
In the context of this simplified model, we consider only 
pair production of the (lightest) stop, $p p \to \tst_1 \tst_1^*$, 
and we assume that the decays $\tst_1 \to t \tz_1$, 
$\tst_1 \to t \tz_2$, and $\tst_1 \to b \tw_1$
occur in the ratio $1:1:2$.
\begin{table}\centering 
\begin{tabular}{lc}
\hline
parameter & value \\
\hline
$m_{\tst_1}$&       $2750.0$ \\
$m_{\tw_1}$ &       $150.0$ \\
$m_{\tz_2}$ &       $150.0$ \\
$m_{\tz_1}$ &       $149.0$ \\
$m_t$      &        $173.2$ \\
\hline
\end{tabular}
\caption{Particle masses in GeV units in a natural SUSY~\cite{rns}
simplified model used to obtain
the optimized cuts~(\ref{stop_cuts_final}) described in the text.}
\label{tab:stop_bm}
\end{table}

\begin{figure}[tbp]
\begin{center}
\includegraphics[width=\lfig,clip]{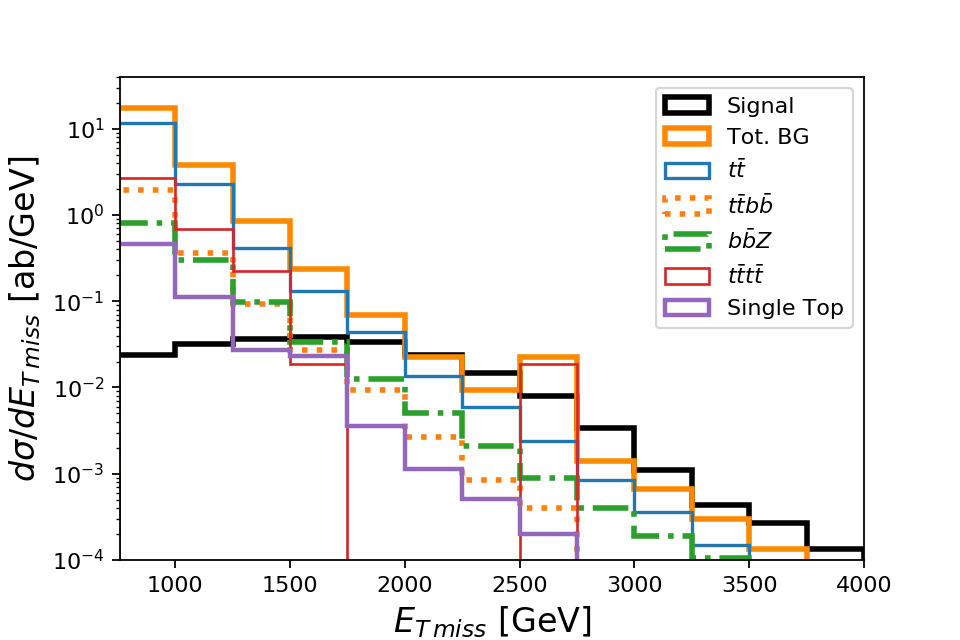}
\caption{
Distribution of $\eslt$ after initial cuts (\ref{stop_cuts_1})
for a $2.75$ TeV stop, as well as for the most
relevant backgrounds, namely $t\bar{t}$, $t\bar{t}b\bar{b}$, $b\bar{b}Z$, $t\bar{t}t\bar{t}$
and single top.
\label{fig:stop_et_missing}}
\end{center}
\end{figure}

We use the stop benchmark point that we have adopted to design cuts to
separate the SUSY signal from SM backgrounds, showing the relevant
distributions at each step.  We begin with the following relatively
basic cuts: 
\bea n_b & \ge & 2, \nonumber\\ 
n_\ell & = & 0, \nonumber\\
\eslt & >& {\rm max}(750\ {\rm GeV}, 0.2 M_{\rm{eff}}),\nonumber\\
E_T(j_1)& > & 500\ {\rm GeV}, \label{stop_cuts_1}\\ 
E_T(j_2)& > & 300\ {\rm GeV}, \nonumber \\ 
S_T &>&0.1 , \nonumber \\ \nonumber \eea 
where
$n_b$ is the number of $b$-tagged jets, $n_\ell$ is the number of
isolated leptons\footnote{ Our isolation criteria
(c.f. Sec.~\ref{subsec:simulation_procedure}) are rather stringent, so
demanding $n_\ell = 0$ removes fewer signal and background events than one
naively might expect.  However, the main point of this lepton veto is to reduce
$t\bar{t}$ backgrounds, which will ultimately be reduced to $\lesssim
0.1$ ab in both our stop and gluino analyses anyway, so making a more stringent
lepton veto by loosening the isolation criteria is unlikely to have a
greater effect in reducing backgrounds than in reducing signals.  } ($e$
or $\mu$), 
$M_{\rm eff}$ is the scalar sum of the $E_T$ of the (up to) four leading
jets and $\eslt$, and $S_T$ is the transverse sphericity, which is
sphericity\footnote{See, {\it e.g.}, {\it Collider Physics}, V. Barger
and R. J. N. Phillips (Addison Wesley, 1987) for the definition of
sphericity.}  defined using only transverse quantities.  We display the
$\eslt$ distribution after the cuts (\ref{stop_cuts_1}) in
Fig.~\ref{fig:stop_et_missing} for the signal benchmark point and for
the backgrounds considered.
As expected, we see that the signal to background ratio (S/B) becomes larger
with increasing $\eslt$.  Nonetheless, to maximize the reach
we choose cuts to preserve
as much signal as possible.  Since the signal $\eslt$ distribution for
our benchmark point, with its relatively heavy stop, attains a broad peak at
$\eslt \sim 1500$~GeV, we require 
\bea
\eslt > 1500~ {\rm GeV}. \label{stop_cuts_2}
\eea
At this stage, $t\bar{t}$ remains the largest contributor to the
background. 

\begin{figure}[tbp]
\begin{center}
\includegraphics[width=\lfig,clip]{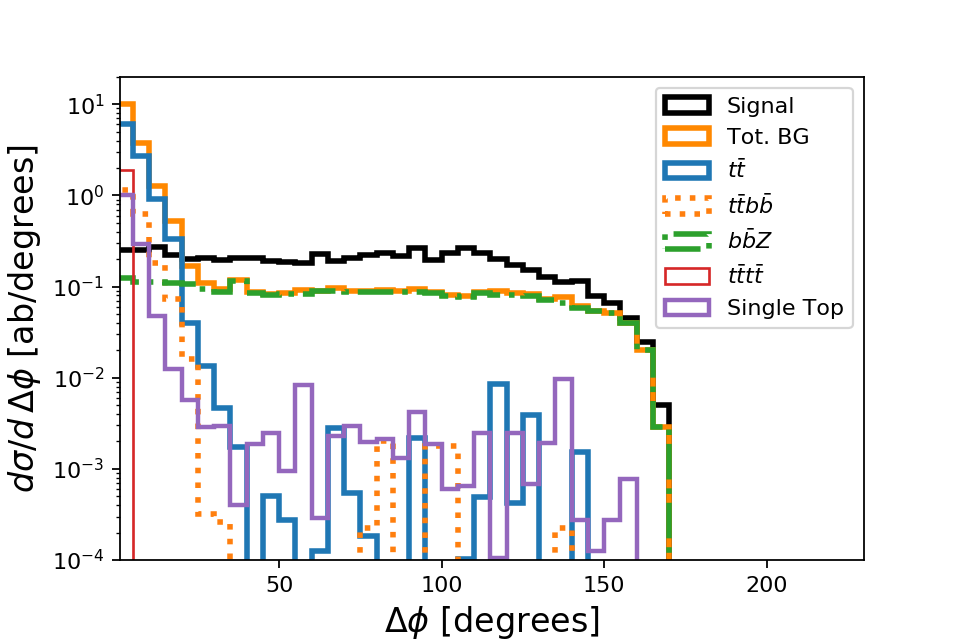}
\caption{
Distribution of $\Delta \phi$ after initial cuts (\ref{stop_cuts_1}) and the subsequent
harder $\eslt$ cut (\ref{stop_cuts_2})
for a $2.75$ TeV stop, as well as for the most relevant backgrounds.
\label{fig:stop_delta_phi}}
\end{center}
\end{figure}

With the more stringent $\eslt$ cut (\ref{stop_cuts_2}) we turn our
attention to other observables, first focusing on cuts to suppress the
dominant $t\bar{t}$ background. Toward this end, we recognize that
$t\bar{t}$ production is most likely to yield events with $\eslt \gg
m_t$ if the tops are produced with large transverse momentum, with one
top decaying semi-leptonically and the other hadronically. In this case,
because the top decay products are strongly boosted, the neutrino from
the semi-leptonically decaying top (which gives the bulk of the $\eslt$)
is likely to be aligned with the $b$-parton jet from the same top,
regardless of whether or not it is tagged as a $b$-jet. This suggests
that we examine the distribution of of $\Delta \phi(\eslt,\ {\rm
nearest~jet})$, the minimum angle between the transverse momenta of any
jet and the $\eslt$ vector in the transverse plane, which we shall
henceforth term $\Delta \phi$. While we expect the $t\bar{t}$ background
(and to a large extent also $t\bar{t}b\bar{b}$ background) to peak at small
$\Delta\phi$, we do not expect this to be the case for the signal since
the higgsino directions are not particularly correlated with those of
the other daughters of the stop.  This
distribution is shown in Fig.~\ref{fig:stop_delta_phi}.

We indeed see that removing events with small $\Delta \phi$ values
~\cite{us-14,atlasanal} will significantly reduce the background with
little effect on the signal.  
We therefore require an additional analysis
cut, 
\bea \Delta \phi > 30~{\rm degrees}. \label{stop_cuts_3} 
\eea 
We see that this cut reduces the background to be smaller than the signal
for the benchmark point and that after this cut 95\% of the SM
background is due to $b\bar{b}Z$ production. To further optimize the signal, we
investigate the distribution of the $E_T$ of the leading jet for
signal and background in Fig.~\ref{fig:stop_leading_jet_pt}. We show
only the total backgound in this and in the following figures. 
\begin{figure}[tbp]
\begin{center}
\includegraphics[width=\lfig,clip]{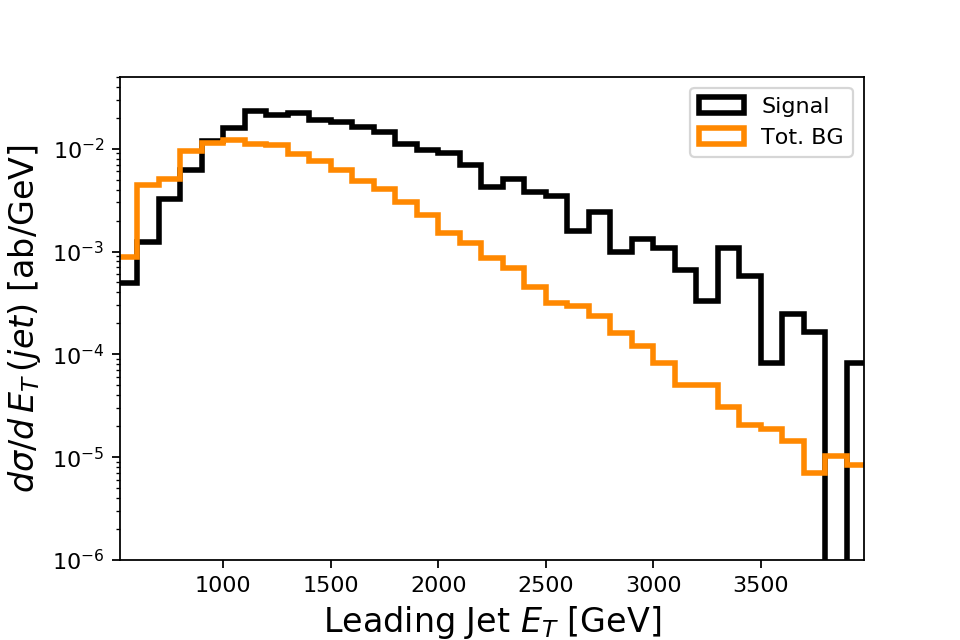}
\caption{ Distribution of leading jet $E_T$ for the signal and 
for the sum of the backgrounds  after initial cuts
(\ref{stop_cuts_1}), the subsequent harder $\eslt$ cut
(\ref{stop_cuts_2}), and the $\Delta \phi$ cut (\ref{stop_cuts_3}).
\label{fig:stop_leading_jet_pt}}
\end{center}
\end{figure}
We observe that the background distribution is peaked at somewhat lower
values than the signal distribution, suggesting that the additional cut,
\bea E_T(j_1) > 1000~{\rm GeV} \label{stop_cuts_4}, 
\eea 
 will be useful.
Finally, after imposing cut (\ref{stop_cuts_4}), we see -- 
from the $E_T$ distribution of the second leading jet for signal and background shown
in Fig.~\ref{fig:stop_second_jet_pt} -- 
that requiring
\bea E_T(j_2) > 600~{\rm GeV}, \label{stop_cuts_5} 
\eea
will further enhance the signal relative to the background.  
\begin{figure}[tbp]
\begin{center}
\includegraphics[width=\lfig,clip]{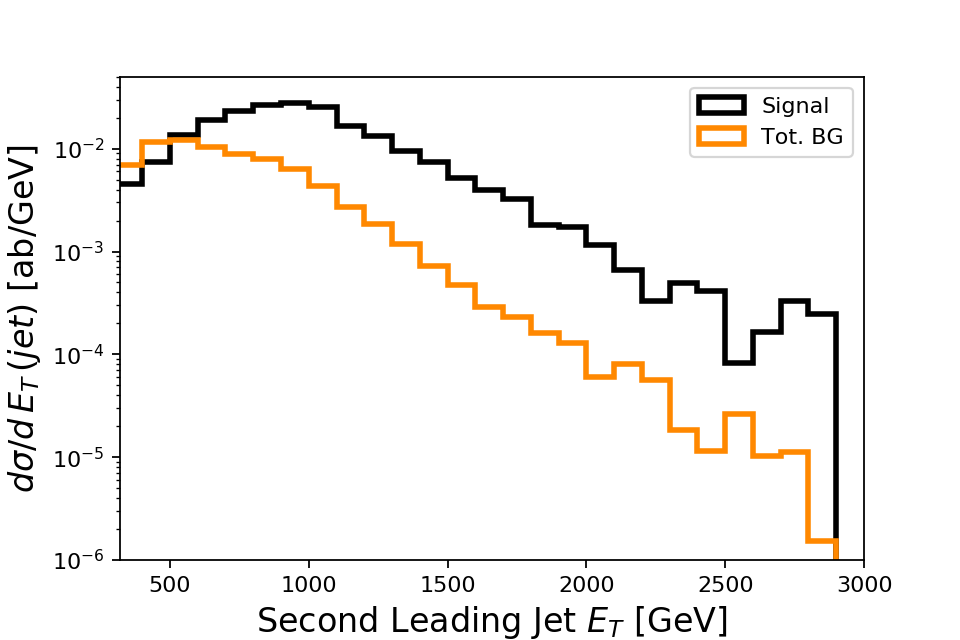}
\caption{ Distribution of second leading jet $E_T$ after initial cuts
(\ref{stop_cuts_1}) and the subsequent harder $\eslt$ cut
(\ref{stop_cuts_2}), $\Delta \phi$ cut (\ref{stop_cuts_3}), and leading
jet $E_T$ cut (\ref{stop_cuts_4}).
\label{fig:stop_second_jet_pt}}
\end{center}
\end{figure}
Cuts (\ref{stop_cuts_1} - \ref{stop_cuts_5}) 
together constitute our analysis cuts for the stop signal at LHC33. 
Combining these cuts, we obtain
\bea
n_b & \ge & 2, \nonumber\\
n_\ell & = & 0, \nonumber\\
\eslt & >& {\rm max}(1500\ {\rm GeV},0.2 M_{\mathrm{eff}}),\nonumber\\
E_T(j_1)& > & 1000\ {\rm GeV},  \label{stop_cuts_final}\\
E_T(j_2)& > & 600\ {\rm GeV}, \nonumber \\
S_T &>&0.1 , \nonumber \\
\Delta \phi & > & 30\ {\rm degrees}. \nonumber \\ \nonumber
\eea
After these cuts, the total background cross section is $4.6$ ab,
as compared to a signal cross section of $18.9$~ab
for the ($2.75$ TeV stop) benchmark model.  
Over 95\% of this background is from $b\bar{b}Z$; the cuts have
effectively rejected the other backgrounds, in particular $t\bar{t}$.
While we will not attempt to optimize our cuts further,
as we wish our results to be robust with respect to the details of
our simulation, this does suggest that a more aggressive 
analysis to separate the $b\bar{b}Z$ background from the 
stop production signal may be possible.

\subsection{Gluino Analysis}

\label{subsec:gluino_analysis}

Next, we turn to the determination of the optimal cuts for gluino
discovery. We had seen in a previous study~\cite{jamieplb} that the
LHC33 reach for gluinos extends beyond $\sim 5$~TeV. Here, we attempt
to choose the gluino analysis cuts to better optimize the reach.
Toward this end, we adopt a SUSY benchmark point in the natural NUHM2
model~\cite{rns} again with $\mu=150$~GeV, but $m_{\tg} \sim 5.4$~TeV,
so the wino mass is 2~TeV, and the bino mass $\sim 1.1$~TeV.  Other
squarks and sleptons are essentially decoupled.  Although from
Fig.~\ref{fig:total_xsec_33_14_TeV} we expect only 30-40 gluino pair
events per 1~ab$^{-1}$ at LHC33, we expect that these events will
efficiently pass stringent selection cuts to separate them from SM
backgrounds with higher efficiency than stop events. This is partly
because gluino events are harder, and partly because gluino pair events
contain four (rather than two) hard $b$-parton jets that we can use to
reduce the SM background. 
We will
focus our attention on the case where the gluino decays exclusively
via $\tg \to \tst_1 \bar{t}, \tst_1^\ast t$.
The backgrounds considered are the same as those examined
in our stop discovery analyses (Sec.~\ref{subsec:stop_analysis}).

\begin{figure}[tbp]
\begin{center}
\includegraphics[width=\lfig,clip]{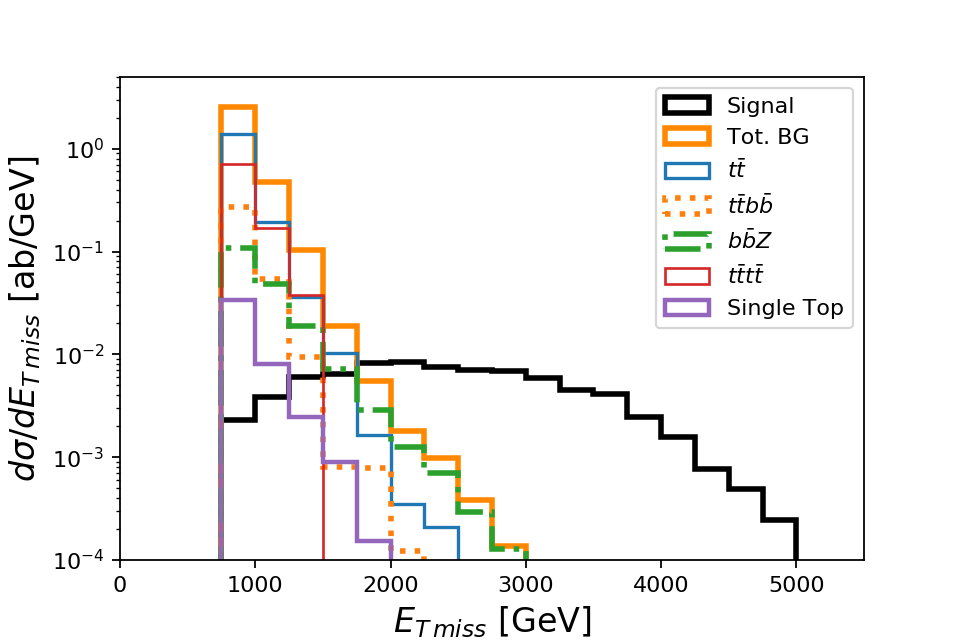}
\caption{ Distribution of $\eslt$ after initial cuts
(\ref{gluino_cuts_1}) for the gluino benchmark point in
Tab.~\ref{tab:bm}, as well as for the most relevant backgrounds, namely
$t\bar{t}$, $t\bar{t}b\bar{b}$, $b\bar{b}Z$, $t\bar{t}t\bar{t}$ and
single top.
\label{fig:gluino_missing_et}}
\end{center}
\end{figure}

\begin{figure}[tbp]
\begin{center}
\includegraphics[width=\lfig,clip]{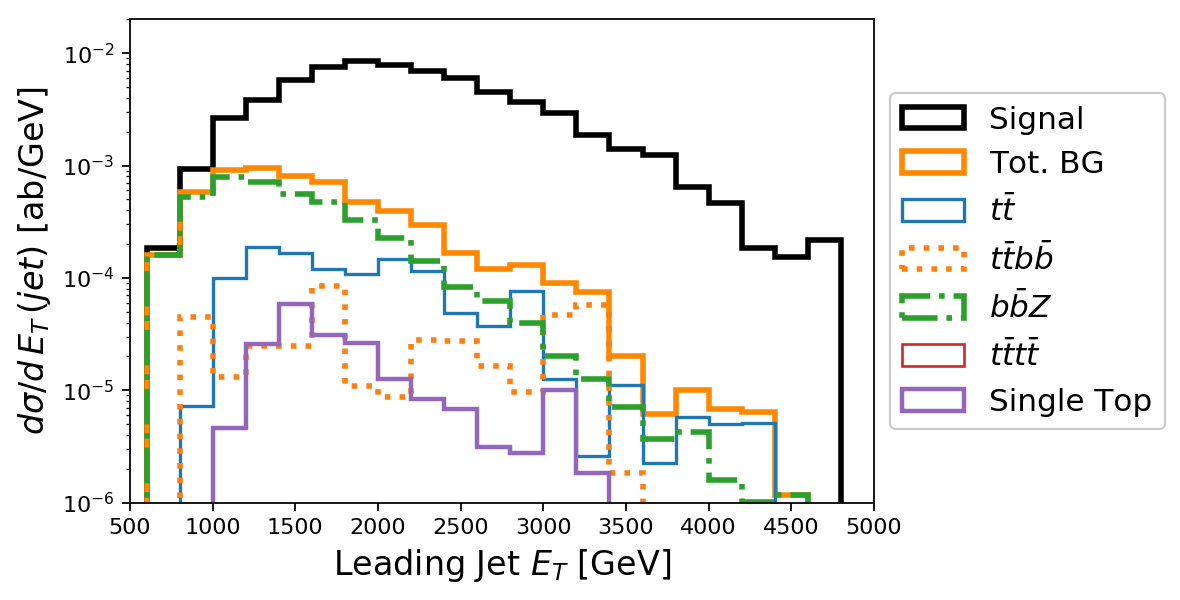}
\caption{
Distribution of leading jet $E_T$ after initial cuts (\ref{gluino_cuts_1}) and the subsequent
harder $\eslt$ cut (\ref{gluino_cuts_2}) for the gluino signal
as well as for the most relevant backgrounds.
\label{fig:gluino_leading_jet_pt}}
\end{center}
\end{figure}

\begin{figure}[tbp]
\begin{center}
\includegraphics[width=\lfig,clip]{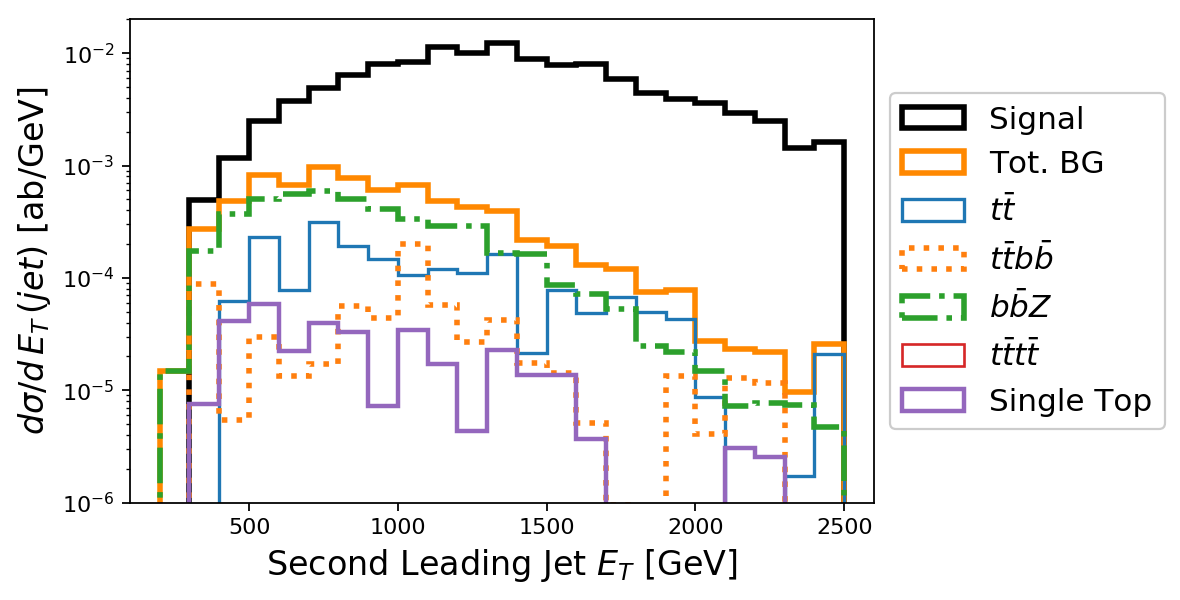}
\caption{
Distribution of second leading jet $E_T$ after initial cuts (\ref{gluino_cuts_1}),
the subsequent harder $\eslt$ cut (\ref{gluino_cuts_2}), and
the $E_T(j_1)$ cut (\ref{gluino_cuts_3})
as well as for the most relevant backgrounds.
\label{fig:gluino_second_jet_pt}}
\end{center}
\end{figure}

The parameters and relevant mass spectrum for the SUSY benchmark point
that we use to develop our cuts for a gluino search at LHC33 are 
shown in Table~\ref{tab:bm}.
\begin{table}\centering
\begin{tabular}{lc}
\hline
parameter & value \\
\hline
$m_0$      & 11750  \\
$m_{1/2}$   & 2350  \\
$A_0$      & -18800  \\
$\tan\beta$&  10  \\
$\mu$      & 150 \\
$m_A$      &  2350 \\
\hline
$m_{\tg}$   &       $5379.0$   \\
$m_{\tst_1}$&       $4257.8$  \\
$m_{\tst_2}$&       $8947.3$ \\
$m_{\tb_1}$ &       $9010.8$ \\
$m_{\tw_2}$ &       $2020.4$  \\
$m_{\tw_1}$ &       $161.6$ \\
$m_{\tz_4}$ &       $2052.0$  \\
$m_{\tz_3}$ &       $1085.6$  \\
$m_{\tz_2}$ &       $157.1$ \\
$m_{\tz_1}$ &       $152.5$ \\
$m_h$      &        $126.7$ \\
\hline
\end{tabular}
\caption{NUHM2 input parameters and some superpartner masses in GeV
units for the {\it radiatively-driven natural SUSY}~\cite{rns} benchmark
point used to develop the cuts~(\ref{gluino_cuts_final}) used in our
gluino discovery analysis.  We take $m_t=173.2$~GeV.}
\label{tab:bm}
\end{table}
We now develop our cuts for gluino discovery, again showing
the relevant distributions at each step.
We begin with the following relatively basic cuts:
\bea
n_b & \ge & 2, \nonumber\\
n_\ell & = & 0, \nonumber\\
\eslt & >& {\rm max}(750\ {\rm GeV},0.2 M_{\mathrm{eff}}),\nonumber\\
n_j     & \ge & 4  \label{gluino_cuts_1} \\
E_T(j_i)& > & 200\ {\rm GeV}, \ (i=1-4) \nonumber\\
S_T &>&0.1 , \nonumber \\
\Delta \phi & > & 10\ {\rm degrees}, \nonumber
\nonumber
\eea
where $n_j$ is the number of jets (with $E_T > 50$ GeV and $|\eta_j| <
3.0$ as noted above).
We display the $\eslt$ distribution after these cuts (\ref{gluino_cuts_1})
in Fig.~\ref{fig:gluino_missing_et} for the gluino signal benchmark point
as well as for the various backgrounds.
Unsurprisingly, S/B increases with $\eslt$.  As before, in the interest
of preserving as much signal as possible, we choose a relatively
conservative cut, keeping in mind that we can rely on other variables to
further improve the signal to background ratio.  Specifically we
require the additional cut,
\bea \eslt > 1900~{\rm GeV}. \label{gluino_cuts_2} 
\eea 
Next, we turn our attention to the $E_T$ distributions of the leading
jets. In
Fig.~\ref{fig:gluino_leading_jet_pt} we display the distribution of the hardest
jet $E_T$ after cuts~(\ref{gluino_cuts_1}) and
(\ref{gluino_cuts_2}),
based on which we implement the additional cut,
\bea
E_T(j_1)  > 1300~{\rm GeV}. \label{gluino_cuts_3}
\eea
We next show the distribution of the second leading jet $E_T$
after the cuts (\ref{gluino_cuts_1} - \ref{gluino_cuts_3}) have been
applied in Fig.~\ref{fig:gluino_second_jet_pt}.
We impose the additional requirement,
\bea
E_T(j_2)  > 900~ {\rm GeV}. \label{gluino_cuts_4}
\eea
We see that the signal, even for this rather heavy gluino benchmark
point, dominates the background. Nevertheless, we checked the
distributions of the third and fourth leading jets and concluded that 
cuts on the $E_T$ (or other properties) of these jets 
did not significantly help with the extraction of the
signal. We do not display these distributions in the interests of brevity.

Thus, the cuts we use in our
gluino discovery analysis at LHC33 are
\bea
n_b & \ge & 2, \nonumber\\
n_\ell & = & 0, \nonumber\\
\eslt & >& {\rm max}(1900\ {\rm GeV},0.2 M_{\mathrm{eff}}),\nonumber\\
n_j    & \ge & 4 \nonumber  \\
E_T(j_1)& > & 1300\ {\rm GeV}, \label{gluino_cuts_final}\\
E_T(j_2)& > & 900\ {\rm GeV}, \nonumber\\
E_T(j_3)& > & 200\ {\rm GeV}, \nonumber\\
E_T(j_4)& > & 200\ {\rm GeV}, \nonumber\\
S_T &>&0.1 , \nonumber \\ \Delta \phi & > & 10\ {\rm degrees}. \nonumber
\nonumber \eea After these cuts the total background cross section is
$0.35$~ab as compared to a signal cross section of $10.4$~ab
for the benchmark model.  
About 55\% of this background arises from $b\bar{b}Z$
production, around 30\% from $t\bar{t}$ production, and $\gtrsim 10$\%
from $t\bar{t}b\bar{b}$ production, with negligible contributions from other
sources.

\section{Stop and Gluino Reach at LHC33}
\label{sec:reach}

We now use the optimized cuts designed in Sec.~\ref{sec:event_selection}
to project discovery reaches and exclusion limits for stops
and gluinos at LHC33.  Our projections for stop reach, which are
presented in Sec.~\ref{sec:stop_reach}, are in the context of the
simplified model with light higginos used to develop the cuts.  
In establishing reach, we will vary the stop mass while leaving the
higgsino masses fixed at around 150~GeV and take the branching fractions
for the stop decays $\tst_1 \to t\tz_1, t\tz_2$ to be 25\% each and the branching fraction for 
$\tst_1\to b\tw_1 = 50$\% (as expected if the stop dominantly decays via
Yukawa couplings). As we discussed earlier, this is representative of
models with light higgsinos and relatively heavy gauginos. We fix the
higgsino masses by choosing $\mu= 150$~GeV. We expect that our results
are insensitive to the precise value as long as $m_{\tst_1}-m_t \gg
|\mu|$.

The corresponding discovery contours and limits on the gluino mass
are obtained in the context of the same simplified model, but (obviously)
with the addition of a gluino, satisfying $m_{\tg} > m_{\tst_1}$.
We expect this will capture the gluino reach in a wide variety of
well-motivated light higgsino models where the gluino decays via
$\tg\to t\tst_1$, with the stop decaying to higgsinos as detailed above.

\subsection{Stop Reach}
\label{sec:stop_reach}

\begin{figure}[tbp]
\begin{center}
\includegraphics[width=\lfig,clip]{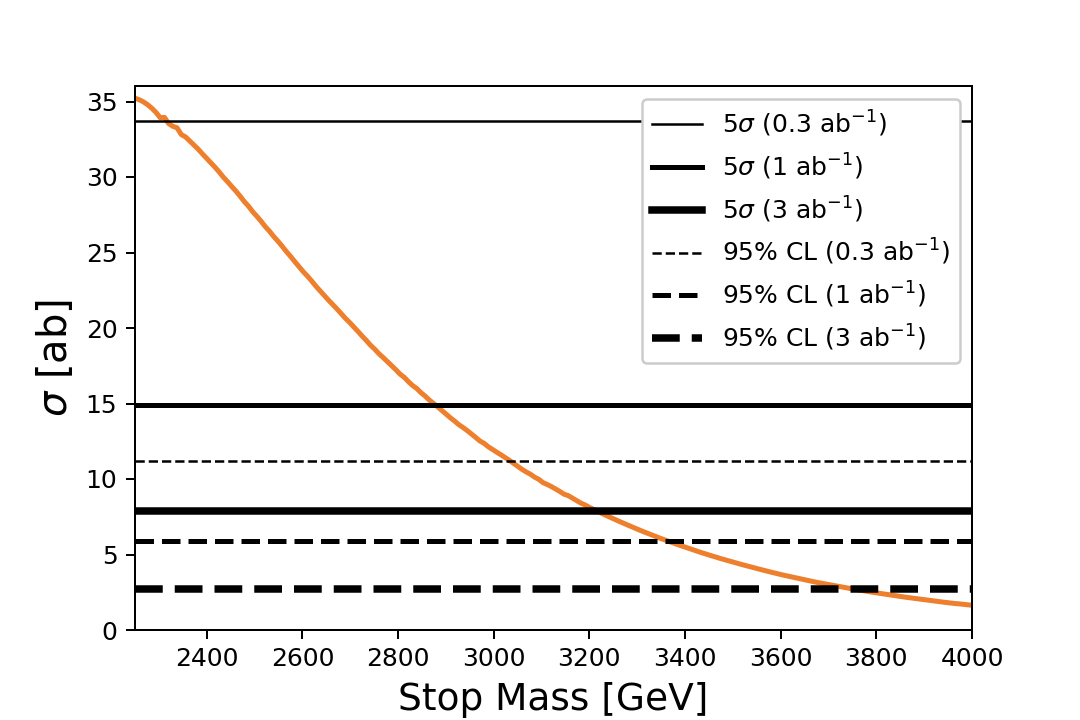}
\caption{
The stop signal cross section after the optimized analyses
cuts (\ref{stop_cuts_final})
  described in the text. The solid horizontal lines
  show the minimum cross section for which an upward Poisson fluctuation of
  the SM background, 
  occurs with a Gaussian probability
  corresponding to $5\sigma$, for integrated luminosities of 0.3, 1 and
  3~ab$^{-1}$ at LHC33.
  The dashed horizontal lines give the signal cross section 
  which is excluded at 95\% for the same choices of 
  integrated luminosity.
\label{fig:stop-reach}}
\end{center}
\end{figure}

In Fig.~\ref{fig:stop-reach}, we show the stop pair production cross
section after the stop analysis cuts (\ref{stop_cuts_final}).
Specifically, we determine the efficiency with which stop signal events
pass the analysis cuts for stop masses of $2250$, $2500$, $2750$,
$3000$, $3250$, $3500$, and $4000$ GeV and interpolate to find the
efficiency for any stop mass.  The total cross section comes from
interpolating the NLO + NLL values from Ref.~\cite{LHC_SUSY}.  As in
Sec.~\ref{subsec:stop_analysis}, we are considering only $p p \to
\tst_1 \tst_1^*$ production.  We are assuming that the lightest two
neutralinos and the lightest chargino are pure higgsinos with the masses
given in Table~\ref{tab:stop_bm} and that stops decay via $\tst_1 \to t
\tz_1$, $\tst_1 \to t \tz_2$, and $\tst_1 \to b \tw_1$ with branching
fractions in the ratio $1:1:2$.

We calculate the signal cross section that will lead to an expected
$5\sigma$ discovery as follows: (a) we take the expected number of
signal plus background events to be the sum of signal and background
cross sections, times the given luminosity, 
rounded to the nearest integer; (b) we find the value of signal cross section
which gives an expected number of signal plus background events for which the Poisson
probability of observing at least that many {\em background only} events is
equal to the $p$-value that corresponds to $5\sigma$, {\it i.e.},
$\approx 2.87 \times 10^{-7}$.  The values of this quantity, {\it i.e.},
$33.7$, $14.9$, and $7.9$ ab for $300$ fb$^{-1}$, $1$ ab$^{-1}$, and $3$
ab$^{-1}$, respectively, are indicated by solid horizontal lines in
Fig.~\ref{fig:stop-reach}.  The discovery reach in this channel is then
the value of the stop mass for which the signal cross section is equal
to these discovery cross sections, {\it i.e.}, the values of stop mass
at which the lighter/ orange cross section curves cross the discovery cross
section lines.  Specifically, we find that the $5\sigma$ mass reach for stop
discovery in this channel is $m_{\tst_1} \approx 2300$, $2900$, and
$3200$ GeV for $300$ fb$^{-1}$, $1$ ab$^{-1}$, and $3$ ab$^{-1}$,
respectively. We regard a signal as observable if we have a
``5$\sigma$'' significance together with
least 5 signal events in the data sample and $S/B > 20\%$. 

We also calculate the signal cross section which is expected to be
ruled out at 95\% confidence level after various integrated luminosities.
Specifically we find the expected number of background events, $B$, 
observed at the given luminosity by multiplying the background 
cross section by the integrated luminosity and rounding to the nearest 
integer.  The signal expected to be ruled out at 95\% confidence level
is then the value of signal for which the Poisson probability of observing
$B$ or fewer events is 5\%.
The values of this quantity, {\it i.e.}, $11.2$, $5.9$, and $2.7$~ab for 
$300$ fb$^{-1}$, $1$ ab$^{-1}$, and $3$ ab$^{-1}$ respectively, 
are indicated by dashed horizontal lines in Fig.~\ref{fig:stop-reach};
from this figure we see that the stop masses that we project
would be excluded at 95\% confidence level if no signal was observed
are indicated by dashed horizontal lines in Fig.~\ref{fig:stop-reach}.
From looking at the values of stop mass for which the 
cross section reaches these values, we find that the values of
the stop mass expected to be ruled out at 95\% are
$3000$, $3400$, and $3800$ GeV for 
$300$ fb$^{-1}$, $1$ ab$^{-1}$, and $3$ ab$^{-1}$ respectively.

These results have been obtained using the value of 4.6~ab for the
background cross section obtained in Sec.~\ref{subsec:stop_analysis}.
Especially since we are considering projections for a future machine and
we are requiring two $b$-tagged jets in all of our analyses, one might
wonder how our results would vary if we considered other
parameterizations of the $b$-tagging efficiencies.  Using the CMS
``medium'' $b$-tagging parameterization~\cite{CMS:2016kkf}, the
background cross section obtained from the analysis cuts
(\ref{stop_cuts_final}) rises to $7.3$ ab and under the CMS ``tight''
$b$-tagging parameterization this cross section falls to $1.8$ ab.
However, the signal cross section is affected in a similar way; using
``medium'' $b$-tagging the signal cross section for the benchmark $2.75$
TeV stop point rises to $23.3$ ab from the $18.9$ ab we obtain with our
default $b$-tagging parameterization (described in
Sec.~\ref{subsec:simulation_procedure}), so using the ``medium''
$b$-tagging parametrization would have minimal effects on our
projections.  Using the ``tight'' $b$-tagging parametrization reduces
the signal cross section at this point to $5.6$ ab.  So our projections
would change more substantially under this scenario, though
re-optimizing our cuts with this $b$-tagging prescription would
presumably lessen
the changes.  Finally, we note that even after scaling up our
backgrounds by a factor of $10$, we would still project 95\% confidence
level exclusions of stop masses below $\sim 3000$ ($3300$) GeV with $1$
($3$) ab$^{-1}$ of integrated luminosity.

\subsection{Gluino Reach}
\label{subsec:gluino_reach}

The efficiency for gluino event detection, and hence the reach for
discovery or exclusion in the gluino mass, also depends on the stop mass.
For this reason, rather than only evaluating the reach along some particular
model line, we examine
the signal cross section in a simplified model identical to the one used
for stop analyses in Secs.~\ref{subsec:stop_analysis} and
\ref{sec:stop_reach}, except that now we also have a gluino, whose mass
is an additional free parameter.  We consider only gluino pair
production, $p p \to \tg \tg$, where each gluino then decays via $\tg
\to \tst_1 \bar{t}$ or $\tg \to \tst_1^\ast t$ with $50$\% probability
for each decay mode.  The stop then decays to higgsinos via $\tst_1 \to
t \tz_1$, $\tst_1 \to t \tz_2$, and $\tst_1 \to b \tw_1$ in the ratio
$1:1:2$.

We determine the efficiency with which generated events pass the
analysis cuts for a number of simplified model points in the $m_{\tg} -
m_{\tst_1}$ plane and interpolate.  In performing this interpolation, we
take the efficiency to go to zero when $m_{\tst_1} + m_t = m_{\tg}$.
This is unduly conservative: as we will see below, there is considerable
sensitivity where the gluino decays via off-shell stops, but these decay
modes are not a part of our simplified model.

\begin{figure}[tbp]
\begin{center}
\includegraphics[width=\lfig,clip]{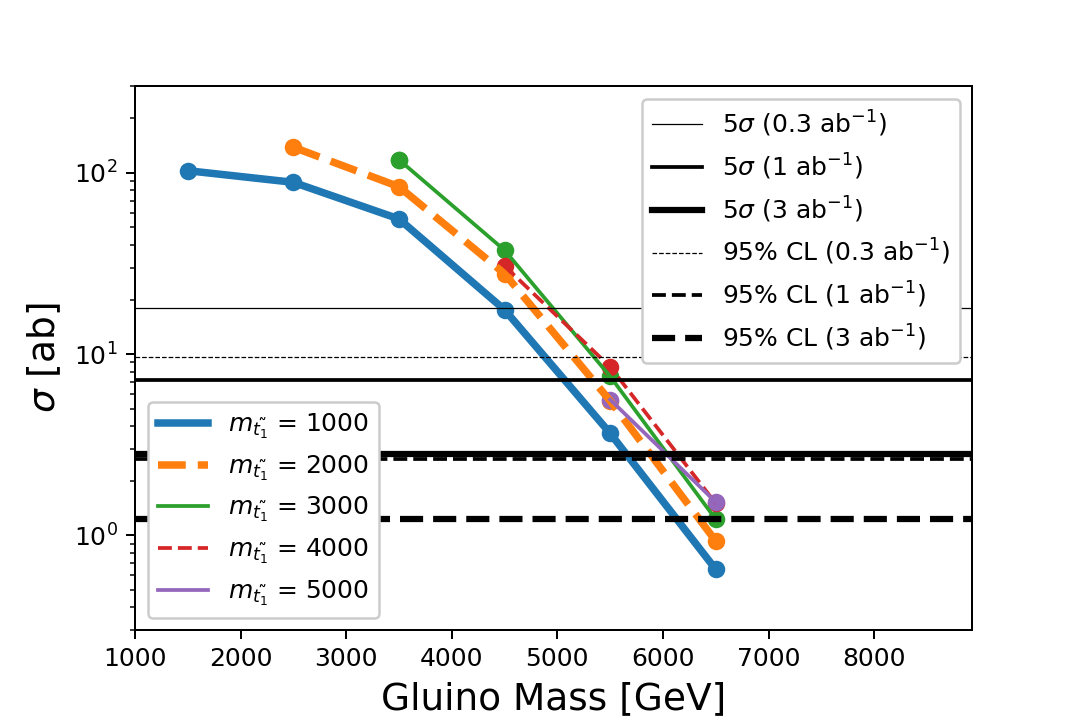}
\caption{ The gluino signal cross section for the optimized analyses
cuts (\ref{gluino_cuts_final}) described in the text for various choices
of stop mass.  The solid horizontal lines show the minimum cross section
for which an upward Poisson fluctuation of the SM background, for which the
cross section is $0.35$ ab, occurs with a Gaussian probability
corresponding to $5\sigma$ for integrated luminosities of 0.3, 1, and
3~ab$^{-1}$ at LHC33.  The dashed horizontal lines give the signal cross
section which is excluded at 95\% for these same values of integrated
luminosity.
\label{fig:gluino-reach}}
\end{center}
\end{figure}

We then multiply our interpolation of efficiency by an interpolation of
the NLO + NLL gluino cross section~\cite{LHC_SUSY}.  The final cross
sections after the analysis cuts~(\ref{gluino_cuts_final}), for different
choices of stop mass, are shown in Fig.~\ref{fig:gluino-reach}.  As in
Fig.~\ref{fig:stop-reach}, the horizontal solid lines represent the
signal cross sections for which we would expect a $5\sigma$ discovery
for the given values of integrated luminosity, and the horizontal dashed
lines represent the signal cross section expected to be excluded at 95\%
confidence level, where these values have been calculated by the same
procedure as in Sec.~\ref{sec:stop_reach}.  We see that with $1$
ab$^{-1}$ the discovery reach in the gluino mass ranges from just over
$5$ TeV (if the stop is light enough to be discovered even at the $14$ TeV LHC) to
around $5.6$ TeV; with $3$ ab$^{-1}$ the discovery reach is over $6$ TeV
for larger values of the stop mass.  Along the model line 
with radiatively-driven naturalness (RNS) studied in
Ref.~\cite{us-14}, the gluino mass discovery reach is $5380$
GeV with $1$ ab$^{-1}$ and $5930$ GeV with $3$ ab$^{-1}$, to be compared
with $\sim 2640$~GeV and 2800~GeV at the HL-LHC.  As in the
case of the stop analyses, using the CMS medium $b$-tagging prescription
raises both signal and background cross sections, while using the tight
prescription will reduce these cross sections.  In neither case would
results be qualitatively changed, and, if we were to re-optimize cuts
for a different $b$-tagging prescription, the overall impact on the
reach would be rather small.

\begin{figure}[tbp]
\begin{center}
\includegraphics[width=\lfig,clip]{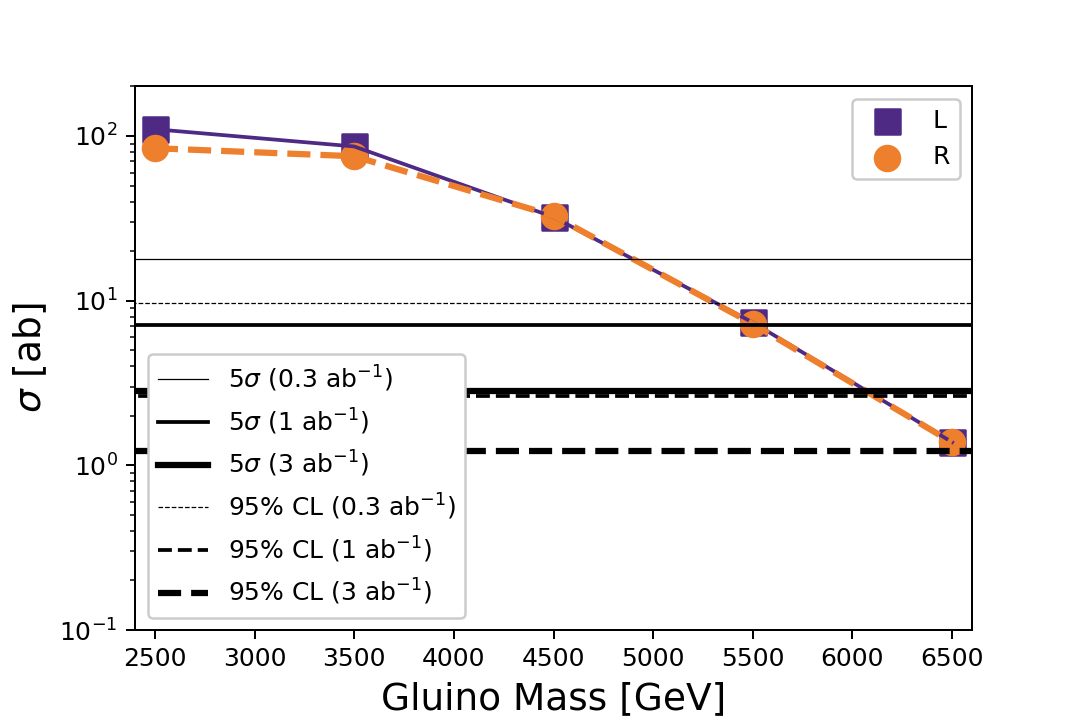}
\caption{ The gluino signal cross section for the optimized analyses
cuts (\ref{gluino_cuts_final}) for the two additional simplified models
in which the gluino decays via three-body decays mediated by $\tst_R$ or
by the left-handed third generation squark doublet 
as described
in the text.  The
solid horizontal lines show the minimum cross section for which an
upward Poisson fluctuation of the SM background, for which the cross section is
$0.35$ ab, occurs with a Gaussian probability corresponding to $5\sigma$
for integrated luminosities of 0.3, 1, and 3~ab$^{-1}$ at LHC33.  
The dashed horizontal lines give the signal
cross section which is excluded at 95\% for several values of integrated
luminosity.
\label{fig:3-body-simplified-reach}}
\end{center}
\end{figure}

The simplified models we have used to obtain the above projections
clearly do not cover all the possibilities.  The first other possibility
is that we have a gluino mass below $m_{\tst_1} + m_t$.  Provided the
light stop (and possibly the light sbottom) mass are much lighter than
the first and second generation squarks, the gluino will decay to a
third generation quark, a third generation antiquark, and a higgsino.
The specific branching ratios depend on the mass of the light stop and
sbottom as well as, {\it e.g.}, the $\tst_L$ component of the light stop.  We,
therefore, consider two additional simplified models to cover such a
possibility.  In one, the lightest stop is purely the right-handed stop
with mass $m_{\tst_1} = 2 m_{\tg}$ and the light sbottom as well as the
heavier stop are decoupled.  In the other, the light stop and sbottom
are both purely left-handed, and $m_{\tsb_1} \simeq m_{\tst_1} = 2
m_{\tg}$.  The discovery reach and exclusion projections for these two
simplified models are shown in Fig.~\ref{fig:3-body-simplified-reach}.
We note that the discovery and exclusion reaches in these scenarios are
(a) very similar to each other and (b) similar to the discovery and
exclusion reaches in the simplified model scenario considered above.  In
particular the $1$ ($3$) ab$^{-1}$ $5\sigma$ discovery reach in these
simplified models is $\approx 5510$ ($6070$) GeV. We conclude that our
gluino reach results are only weakly dependent on the third generation
squark mass as long as we are not extremely close to the kinematic
boundary for two-body gluino decays.

The last possibility for decoupled first/second generation squarks is
that $m_{\tg} > m_{\tst_1} + m_t$, but the gluino also decays to other
third generation 
squarks, such as $\tst_2$.  It is clearly more difficult to
definitively probe this scenario in generality, but we note that if
additional light stops or sbottoms play a prominent role in the most
important decay chains, as we would generally expect, then we would
still expect to be able to observe such models with the cuts given
above with comparable efficiency, so our reach results would not be
qualitatively altered.

\subsection{Distinguishing Stops from Gluinos}
\label{subsec:contamination}

Up to this point our focus has been on the discovery of {\em new
physics} due to the production of stops or gluinos at LHC33: 
{\it i.e.}, in our projections of the stop and gluino reaches, we have only been
concerned whether the SUSY signal can be detected above various SM
backgrounds, regardless of its origin. In this section, we examine whether the
observation of a signal via  either the stop or the gluino search
strategy enables us to unequivocally claim the discovery of the
corresponding particle (assuming a SUSY origin of the signal) or
whether contamination between channels can lead us to
incorrect inferences. 

\begin{figure}[tbp]
\begin{center}
\includegraphics[width=\lfig,clip]{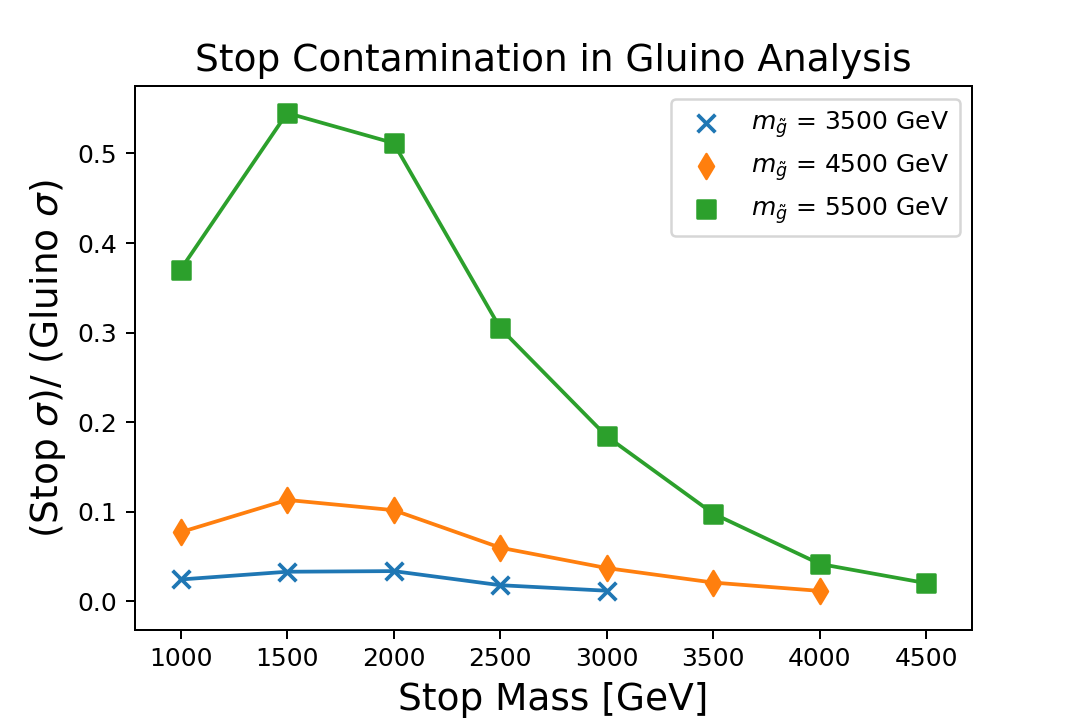}
\caption{The ratio of (i) the cross section for (contamination from) 
stop pair production events at LHC33 the passing the gluino analysis cuts
  (\ref{gluino_cuts_final}) to (ii) the corresponding cross section for
  signal events from gluino pair production, with subsequent gluino decays to ${\tst_1}
  \bar{t}$ or ${\tst_1}^\ast t$, passing the gluino analysis cuts, versus $m_{\tst_1}$, 
  for three values of the signal gluino mass.
\label{fig:stop_contamination_from_gluinos}}
\end{center}
\end{figure}

Toward this end, in Fig.~\ref{fig:stop_contamination_from_gluinos}, we
show the ratio of the cross section from stop production (contamination)
and that for the gluino signal, after the gluino analysis
cuts (\ref{gluino_cuts_final}).  We illustrate this ratio as a function
of $m_{\tst_1}$ for three representative values of the signal gluino
mass.  We see that the stop contamination, though considerably larger than the
SM background of 0.35~ab, is significantly smaller than the gluino
signal, except for values of $m_{\tg}$ approaching the LHC33 reach. The
contamination falls off not only for very heavy stops, for which the
production is kinematically suppressed, but also for relatively light
stops because stop events then have a very low efficiency for passing
the gluino cuts.

\begin{figure}[tbp]
\begin{center}
\includegraphics[width=\lfig,clip]{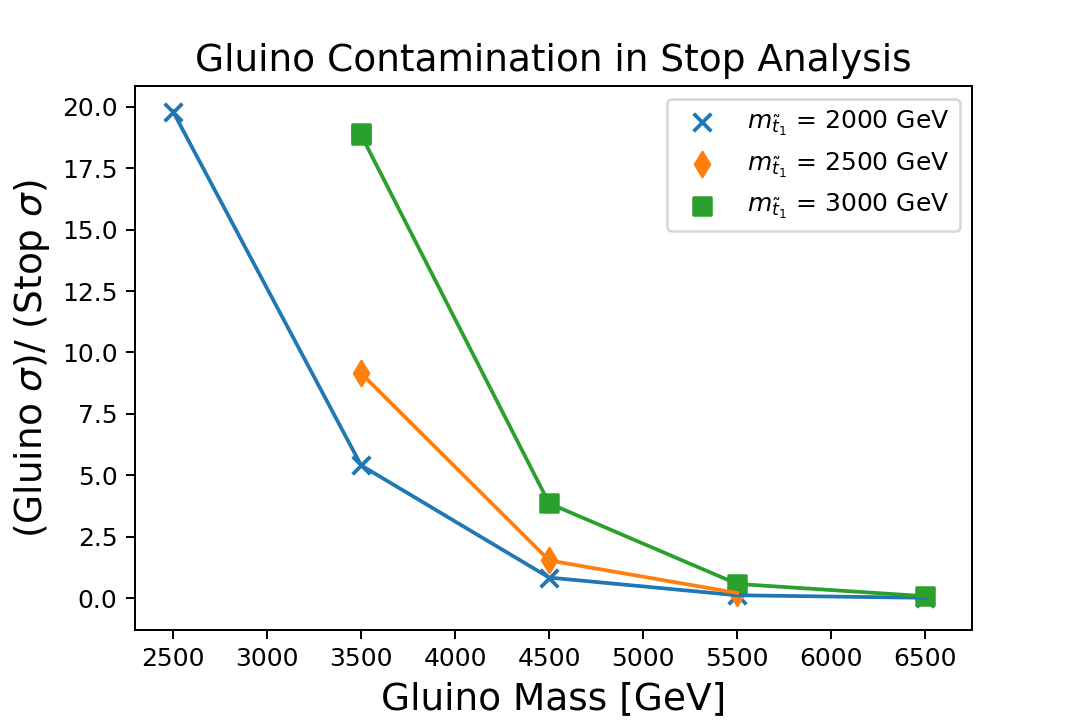}
\caption{
The ratio of (i) the cross section for (contamination from) 
gluino pair production, with subsequent gluino decays to ${\tst_1}
  \bar{t}$ or ${\tst_1}^\ast t$,  for LHC33 events
passing the stop analysis cuts
  (\ref{stop_cuts_final}) to (ii) 
  the corresponding cross section for
  signal events from
  stop pair production passing the same analysis cuts, versus $m_{\tg}$, for three values of
  the signal stop mass.
\label{fig:gluino_contamination_in_stop_decays}}
\end{center}
\end{figure}

The opposite situation -- contamination of the stop signal from gluino
production events -- is illustrated in
Fig.~\ref{fig:gluino_contamination_in_stop_decays} as a function of
$m_{\tg}$.
We see that the situation is quite different in that, unless the gluino
is very heavy, the stop sample is typically dominated by gluino
production events.  This should not be surprising because gluino pair
production, which has a much higher cross section, also results in the
production of stop pairs.  Since gluino production can contaminate
the stop sample with a large rate, it is clear that a discovery of new
physics via the stop selection cuts (\ref{stop_cuts_final}) would not
unequivocally signal direct stop pair production, even if we assume the
new physics has a SUSY origin~\cite{kadala}.

Before turning to the examination of whether stop and gluino pair
production mechanisms can be disentangled from one another, we stress
that the contamination of the stop sample by gluino events, or {\em
vice-versa}, does not negatively affect the extraction of the new physics
reach that we have discussed in the last section. If anything, this reach
will be enhanced because events from gluino production will also
contribute to the signal in the stop analysis, 
and, to a much smaller degree, the other way around.  

Turning to an examination of different ways to determine whether a SUSY
signal arises from gluino or direct stop pair production, we note that gluino
events contain two additional hard tops in addition to the stop pair.
Thus, (a)~we expect the $b$-multiplicity to be greater in the gluino pair
production case, and (b) we typically expect additional {\em very hard
jets} (and perhaps also leptons) in the gluino pair production case.
%
\begin{figure}[tbp]
\begin{center}
\includegraphics[width=\lfig,clip]{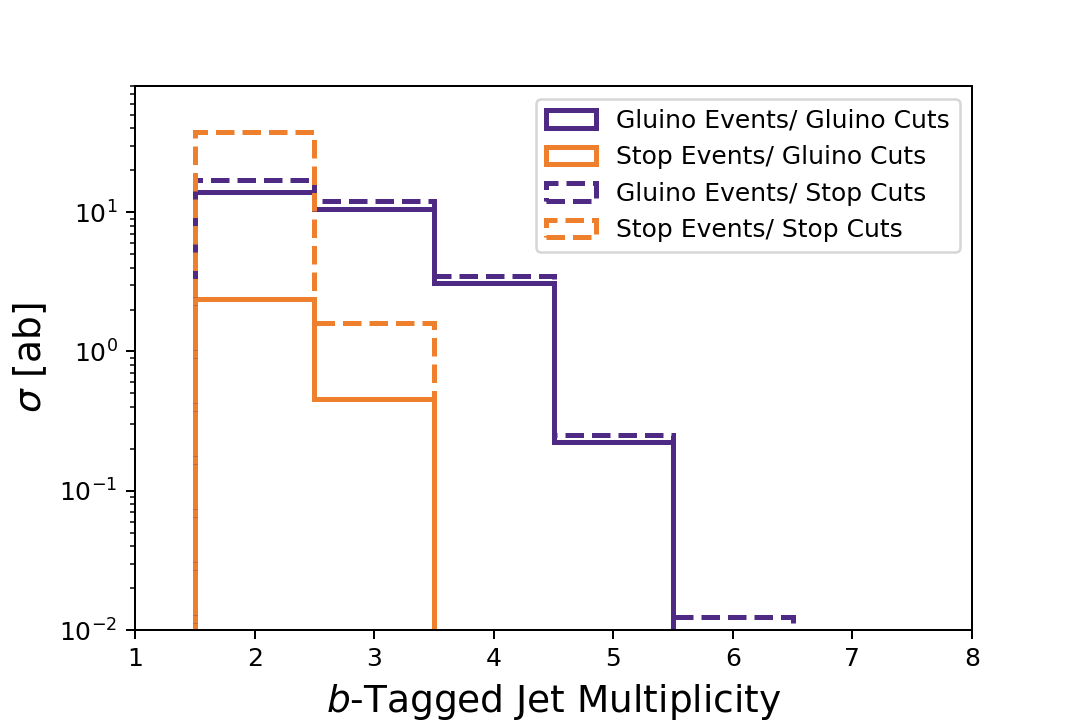}
\caption{ The distribution of multiplicity of tagged $b$-jets in events
from gluino pair production (darker/ purple) and from stop pair
production (lighter/ orange) in a simplified model with $m_{\tg} = 4.5$ TeV
and $m_{\tst_1} = 2$ TeV after the stop analysis cuts (dashed histogram)
(\ref{stop_cuts_final}) and after the gluino analysis cuts (solid
histogram) (\ref{gluino_cuts_final}).
\label{fig:contamination_bjet_mult}}
\end{center}
\end{figure}
%
To illustrate these differences, we consider a simplified model point
with a gluino mass of $4.5$ TeV and a stop mass of $2$ TeV.  We show the
tagged $b$-jet multiplicity distribution for events which pass the stop
[gluino] selection cuts (\ref{stop_cuts_final})
[(\ref{gluino_cuts_final})] in Fig.~\ref{fig:contamination_bjet_mult} as
the dashed [solid] histograms.  Distributions from stop production are
shown by lighter/ orange histograms, while those from gluino production are
shown by the darker/ purple histograms. We see that events with $n_b\ge 3$,
which constitute about 10 - 15\% (which would require ab$^{-1}$ scale
integrated luminosities for observable rates depending on $m_{\tg}$) of
the gluino sample, can only come from gluino production.

\begin{figure}[tbp]
\begin{center}
\includegraphics[width=\lfig,clip]{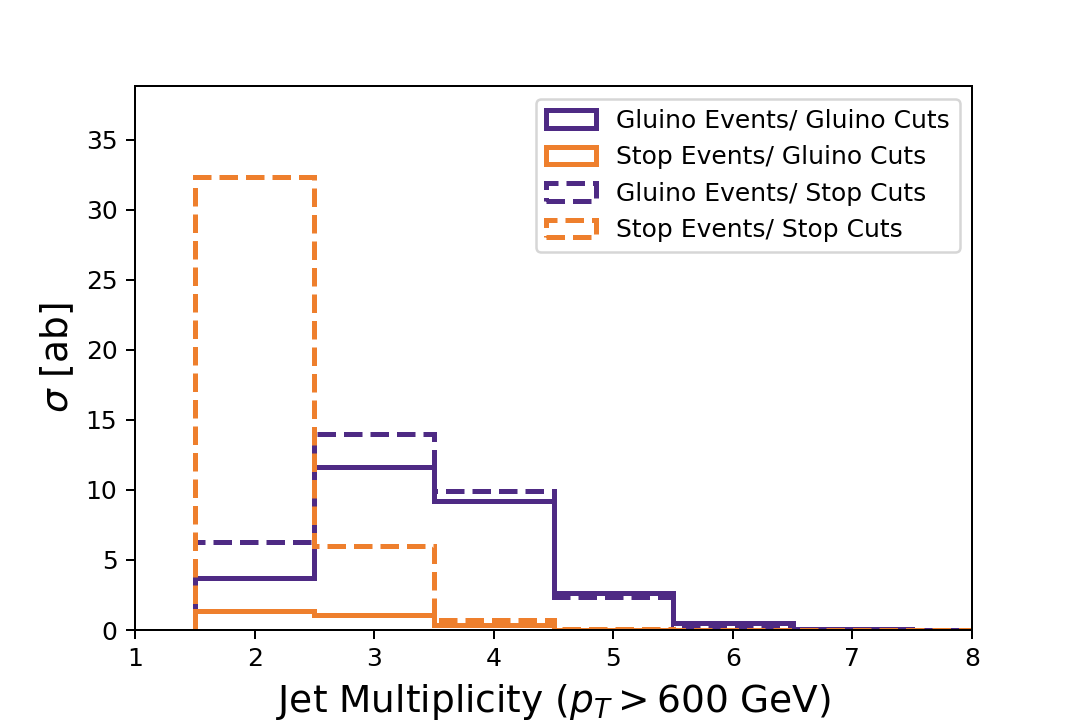}
\caption{ 
The distribution of jet multiplicity for jets with
$E_{Tj}> 600$~GeV in events
from gluino pair production (darker/ purple) and from stop pair
production (lighter/ orange)
in a simplified model with $m_{\tg} = 4.5$ TeV and
$m_{\tst_1} = 2$ TeV after the stop analysis cuts (dashed histogram) 
(\ref{stop_cuts_final}) and after the gluino analysis cuts (solid histogram)
(\ref{gluino_cuts_final}).
\label{fig:jetmult600}}
\end{center}
\end{figure}

We show the jet multiplicity distribution for jets with $E_T > 600$~GeV
in Fig.~\ref{fig:jetmult600}.\footnote{We have checked that though the
jet multiplicity distributions (with jets being defined as 50~GeV
hadronic clusters) from stop and gluino pair production are also
reasonably separated, these are not as distinct as the multiplicity 
distributions of just
the very hard jets shown in Fig.~\ref{fig:jetmult600}, in part because
of radiated QCD jets, and perhaps also because decay products of very
boosted higgsinos can sometimes form a 50~GeV jet. By focussing on hard jets,
we are more likely to count the jets only from the primary decays of the
stops/gluinos.} We see that, after the stop selection cuts, the bulk of
stop events have just two very hard jets, while exactly the reverse is
true for gluino events. Indeed, we see that this distribution serves to
separate the gluino and stop production mechanisms, except perhaps in
the case that $m_{\tg}$ is unfortunately close to $m_{\tst_1}$.

\begin{figure}[tbp]
\begin{center}
\includegraphics[width=\lfig,clip]{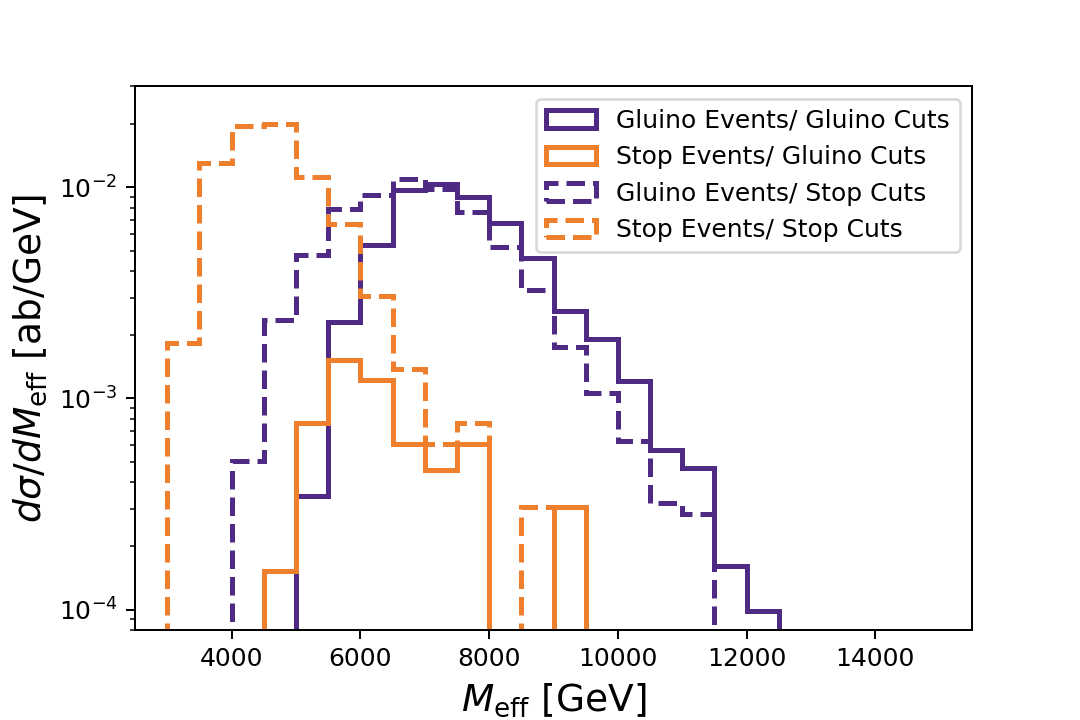}
\caption{The distribution of $M_{\mathrm{eff}}$ 
in events
from gluino pair production (darker/ purple) and from stop pair production
(lighter/ orange)
in a simplified model with $m_{\tg} = 4.5$ TeV and
$m_{\tst_1} = 2$ TeV after the stop analysis cuts (dashed histogram) 
(\ref{stop_cuts_final}) and after the gluino analysis cuts (solid histogram)
(\ref{gluino_cuts_final}).
\label{fig:contamination_meff}}
\end{center}
\end{figure}

\begin{figure}[tbp]
\begin{center}
\includegraphics[width=\lfig,clip]{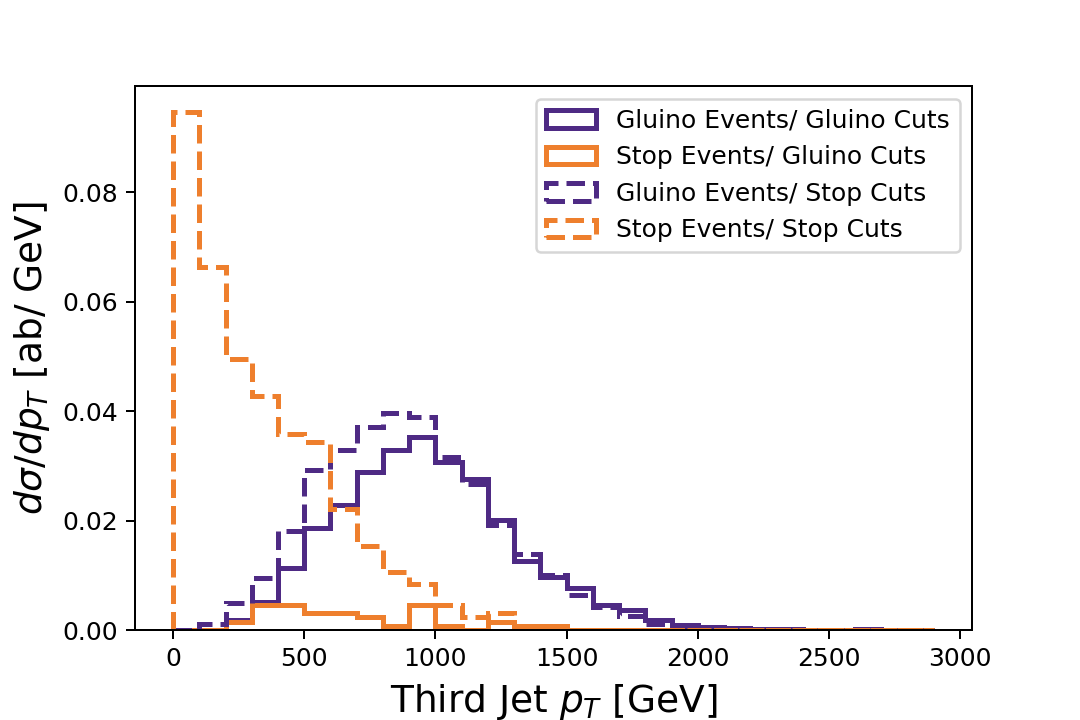}
\caption{The distribution of multiplicity of $E_T(j_3)$ 
in events
from gluino pair production (darker/ purple) and from stop pair
production (lighter/ orange)
in a simplified model with $m_{\tg} = 4.5$ TeV and
$m_{\tst_1} = 2$ TeV after the stop analysis cuts (dashed histogram) 
(\ref{stop_cuts_final}) and after the gluino analysis cuts (solid histogram)
(\ref{gluino_cuts_final}).
\label{fig:contamination_Third_Jet_pT}}
\end{center}
\end{figure}

Unless $m_{\tst_1}$ is relatively close to $m_{\tg}$, further separation
between gluino and stop production events may be possible using kinematic
distributions such as the $M_{\mathrm{eff}}\equiv E_T(j_1)+E_T(j_2)+
E_3(j_3)+E_T(j_4)+\eslt$ distribution shown in
Fig.~\ref{fig:contamination_meff}, or the third jet $E_T$ distribution
shown in Fig.~\ref{fig:contamination_Third_Jet_pT}.
However, if the gluino is just a little heavier than the stop, it is
possible that these distributions may merge, and it may be difficult to
unequivocally tell whether we have stop pair production in addition to
the large gluino signal. Even in this difficult case, progress may be
possible. For instance, it should be possible to extract the gluino
signal and determine its mass from the total rate for gluino events as
proposed in Ref.~\cite{us-14} (the stop contamination is completely
negligible in such a case). This can then be used to project gluino
contamination in the stop sample, enriched using additional hard jet
multiplicity and $n_b$ cuts. A detailed study of this is beyond the
scope of the present analysis.

\section{Implications of LHC33 Reach Results for Natural SUSY Models}
\label{sec:implications}

We have just seen that in models with light higgsinos, experiments at
LHC33 should be able to discover stops with masses up to $2.8 - 3.2$~TeV
and gluinos out to $\sim 5.5 - 6.0$~TeV, assuming that gluinos dominantly
decay via $\tg \to \tst_1 t$ (or via three-body decays to third generation
quarks and higgsinos). In this section, we focus on the
implications of these results for natural SUSY models that have received
much attention over the last several years.

Many authors have argued that SUSY naturalness requires light stops, 
with masses ranging from a few hundred GeV to around 1~TeV,
depending on the degree of fine-tuning that is allowed. It has also been
suggested that the current LHC lower limits on stop masses are
already in tension with naturalness. Implicit in this reasoning is the
unstated assumption that the underlying model parameters are
independent. The arguments that lead to these strong bounds on stop
masses do not apply if the underlying model parameters are correlated,
as will almost certainly be the case once the underlying SUSY-breaking
mechanism is understood. Here, we adopt a more conservative approach to
fine-tuning~\cite{ltr,rns} which recognizes the possibility that model
parameters (usually taken to be independent) might be correlated in the
underlying theory.

Toward this end, we begin with the well-known expression that yields the
measured value of $m_Z$ in terms of the weak scale SUSY Lagrangian
parameters,
\be
\frac{m_Z^2}{2}=\frac{m_{H_d}^2+\Sigma_d^d-(m_{H_u}^2+\Sigma_u^u)\tan^2\beta}{\tan^2\beta
-1}-\mu^2 \simeq -m_{H_u}^2-\Sigma_u^u -\mu^2,  \label{eq:mzs}
\ee
where the last equality is valid for moderate to large values of
$\tan\beta$.  The quantities $\Sigma_u^u$ and $\Sigma_d^d$ in
Eq.~(\ref{eq:mzs}) arise from 1-loop corrections to the scalar potential
(their forms are listed in the Appendix of Ref.~\cite{rns}), $m_{H_u}^2$
and $m_{H_d}^2$ are the soft SUSY-breaking Higgs mass parameters, $\tan
\beta \equiv \langle H_u \rangle / \langle H_d \rangle$ is the ratio of
the Higgs field VEVs, and $\mu$ is the superpotential (SUSY conserving)
Higgs/higgsino mass parameter.  SUSY models requiring large
cancellations between the various terms on the right hand side of
Eq.~(\ref{eq:mzs}) to reproduce the measured value of $m_Z^2$ are
regarded as fine-tuned. With this in mind, we adopt the 
{\it electroweak} fine-tuning measure, $\Delta_{\rm EW}$,  which compares 
the maximum absolute value of each term on the right-hand-side of
Eq.~(\ref{eq:mzs}) to the left-hand-side, $m_Z^2/2$.

The electroweak fine-tuning parameter yields the minimal fine-tuning for
a given sparticle spectrum. We advocate its use for discussions of
fine-tuning because it allows for the possibility that SUSY-breaking
parameters (frequently assumed to be independent in the evaluation of
fine-tuning) might be correlated in the underlying SUSY
theory. Moreover, as we have previously argued, the traditionally used
fine-tuning measure, $\Delta_{\rm BG}$~\cite{bg}, reduces to $\Delta_{\rm
EW}$ once correlations between parameters are correctly incorporated
~\cite{reduces}.  We stress that $\Delta_{\rm BG}$, computed without
parameter correlations being implemented, could easily be two orders of
magnitude larger than $\Delta_{\rm EW}$. For this reason, discarding
models as unnatural because $\Delta_{\rm BG}$ (computed naively) is $\sim
100-1000$ could exclude perfectly viable underlying theories. In this
paper, we conservatively adopt $\Delta_{\rm EW}< 30$ (corresponding to
no more than a part in thirty electroweak fine-tuning) as a criterion
for naturalness of the superpartner spectrum.
The onset of electroweak fine-tuning for values $\Delta_{\rm EW}\agt 20-30$ is 
visually displayed in Refs.~\cite{upper} and~\cite{Baer:2017yqq}. 

\subsection{Radiatively-driven Natural SUSY (RNS)} 
\label{subsec:rns}

We see from Eq.~(\ref{eq:mzs}) that in order to obtain low values of
$\Delta_{\rm EW}$, the weak scale values of $m_{H_u}^2$ and $\mu^2$ must
both have magnitudes not much larger than $m_Z^2$.  At the same time,
LHC search constraints seem to require at least TeV-scale high scale
soft terms. These two facts can be reconciled in models with
radiatively-driven naturalness (radiative natural SUSY or RNS) wherein a
large GUT scale value of $m_{H_u}^2$ is driven to small weak scale
values via renormalization group running.  In the focus point region of
the mSUGRA/CMSSM model, $m_{H_u}^2$ is indeed driven to small negative
values.  However, the FP region is mainly viable for small $A_0$ values
while the rather large Higgs mass $m_h\simeq 125$ GeV requires large
mixing from large $A$ terms.  The main obstacle in the mSUGRA/CMSSM
model to having both naturalness and $m_h\simeq 125$ GeV arises from the
rather unmotivated assumption of scalar mass universality,
$m_{H_u}=m_{H_d}=m_0$, at the GUT scale.  Such universality is not
expected to occur in SUSY GUT models since the Higgs multiplets live in
different GUT representations than the matter multiplets.

This led us to exmine the two-extra-parameter non-universal Higgs
model~\cite{nuhm2} with $m_{H_u}\ne m_{H_d}\ne m_0$ always allows the
freedom to adjust $m_{H_u}^2({\rm GUT})$ to whatever value is needed such that
it is driven to small negative values at the weak scale.\footnote{A
common objection is that then the high scale value of $m_{H_u}^2$ must
be tuned just right to be driven to small negative values. However, if
the proper correlations amongst soft terms are present, then a
generalized focus point mechanism operates such that a large range of
$m_{H_u}^2$ values are all focussed to small negative values at the weak
scale: see Ref.~\cite{guts}.}  In NUHM2, the independent GUT scale
values of $m_{H_u}^2$ and $m_{H_d}^2$ can be traded for weak scale
values of $\mu$ and $m_A$ so that a natural value of $\mu\sim 100-300$
GeV can always be chosen. Then Eq.~(\ref{eq:mzs}) can be used to
determine the weak scale value of $m_{H_u}^2$ which is then driven via
RGEs to determine $m_{H_u}^2({\rm GUT})$. Thus, in NUHM2 (and other SUSY
models which allow for high-scale freedom in the Higgs soft terms),
electroweak naturalness, {\it i.e.}, $\Delta_{\rm EW} < 30$, can be
satisfied if the high scale $\mu$-parameter, which typically does not
evolve very much, is selected to be $\sim m_{weak}$ and the $\Sigma_u^u$
terms are not too large.

The most robust feature of RNS models is that these include four light
higgsinos-like states $\tz_1,\tz_2,\tw_1^\pm$ with masses $\sim
100-300$~GeV because the weak scale value of $\mu^2$ directly enters
Eq.~(\ref{eq:mzs}).\footnote{Here, we implicitly assume that the
superpotential parameter, $\mu$, is the dominant source of the higgsino
mass. A soft SUSY-breaking contribution to the higgsino mass is possible
if there are no additional gauge singlets that couple to higgsinos
~\cite{ross}. There are also extended frameworks with additional TeV
scale fields that show it is theoretically possible to decouple the
higgsino mass from the Higgs boson mass parameter that enters into
Eq.~(\ref{eq:mzs})~\cite{other}.} 
If bino and wino mass parameters are large 
(which is typically the case in models with gaugino mass unification which
obey LHC gluino mass constraints),
the mass splitting between the higgsinos is small, 
so the visible decay products of $\tz_2$ and
$\tw_1^\pm$ are very soft so that signals from electroweak higgsino
pair production are typically buried below SM backgrounds at the LHC. 
We should mention that electroweak naturalness also imposes upper bounds on 
stop and gluino masses in the RNS framework. The bounds on stop
masses arise because these produce large corrections to the Higgs
effective potential and are manifested by large contributions to 
$\Sigma_u^u$ in Eq.~(\ref{eq:mzs}). A low value of $\Delta_{\rm EW}$
requires a large SUSY-breaking trilinear stop parameter comparable to
the (third generation) scalar mass parameter. This simultaneously leads
to sizeable left-right stop mixing and automatically lifts the light
Higgs boson mass to its observed value~\cite{ltr}. The upper limit on
$m_{\tg}$ arises because heavy gluinos result in large radiative
corrections to the stop masses~\cite{sundrum}. We will return to these
stop and gluino limits, which are (only weakly) sensitive to the details
of the underlying RNS model, later in this section.

\subsubsection{The natural  NUHM2 model}\label{nuhm2} 

The NUHM2 model, with two additional parameters $m_{H_u}^2({\rm GUT})$
 and $m_{H_d}^2({\rm GUT})$ {\em vis-\`a-vis} the mSUGRA/CMSSM model, is
the prototypical RNS model. One special feature of this NUHM2 RNS model
is the assumed gaugino mass unification which constrains the wino mass
to be $\sim m_{\tg}/3$ and the bino mass to be $\sim m_{\tg}/6$.  This
has two important consequences:

\begin{enumerate}

\item As the integrated luminosity accumulated at the LHC increases, it
has been shown that wino pair production provides a deeper reach into
parameter space than gluino pair production.  Wino pair production leads
to clean same-sign dilepton events free from jet activity (other than
that from QCD radiation) via same sign diboson (SSdB) production
processes~\cite{lhcltr} arising from $pp\to\tw_2^\pm \tz_4$ with $\tw_2^\pm
\to W^\pm\tz_{1,2}$ and $\tz_4\to W^\pm\tw_1^\mp$.  Since the
naturalness upper limit on $m_{\tg}$ also implies a corresponding upper
limit on the wino mass, this clean signal channel leads to a HL-LHC (3000
fb$^{-1}$) reach to $m_{1/2}\sim 1.2$ TeV, covering nearly all of the
$\Delta_{\rm EW} <30$ region consistent with the observed value of Higgs
boson mass.

\item In the NUHM2 RNS model, the fact that the wino and bino masses
are bounded above implies that
higgsino mass gap, $m_{\tz_2}-m_{\tz_1}$, though small, is always
larger than $\sim 10$~GeV over the entire range of parameters with
$\Delta_{\rm EW}< 30$.  Although electroweak production of higgsinos is
swamped by SM backgrounds due to the small visible energy release in
higgsino decays, higgsino pair production in association with a hard QCD
jet-- for instance $pp\to\tz_1\tz_2+jet$ or $\tw_1\tz_2+jet$ with
$\tz_2\to\tz_1\ell^+\ell^-$-- offers a HL-LHC reach to $\mu\sim 250$
GeV~\cite{llj}.  The detectability of the soft dilepton pair with
$m_{\ell\ell} < m_{\tz_2}-m_{\tz_1}$, which is very sensitive to the
size of the 
neutralino mass gap, plays a critical role in limiting the SM
background.

\end{enumerate} 
A combination of these two signals covers essentially 
the entire
parameter space of the natural NUHM2 model with $\Delta_{\rm EW} < 30$
at the HL-LHC.  Unfortunately, this cannot be said to cover all natural
SUSY models since the assumption of gaugino mass unification, which
played a critical role in our discussion, can be relaxed without
affecting naturalness. We are thus led to examine other well-motivated
natural SUSY models where gaugino masses do not unify at $Q=M_{\rm
GUT}$, and where weak scale wino and bino mass parameters are closer to
the gluino mass.

\subsubsection{The natural NUHM3 model} 

The natural NUHM3 model is an obvious extension of the natural NUHM2 model
where we allow the high scale third generation mass, $m_0(3)$, to be
independent of the first and second generation scalar masses. Gaugino mass
parameters are assumed to be unified at $Q=M_{\rm GUT}$. This framework
is partly motivated because it allows very heavy first and second generation
squarks -- 
which ameliorates the SUSY flavor problem -- while keeping the third
generation in the TeV range which is 
consistent with naturalness considerations. 
From the theory side, it is worth noting that there are 
extra-dimensional top-down models where the third generation feels SUSY
breaking differently from the first two generations because of the
geography of extra dimensions; see Sec.~\ref{subsubsec:miniland} below.
As discussed in Sec.~\ref{subsec:glstop}, the important point is that this
additional freedom allows considerably heavier gluinos than the NUHM2
RNS model.

\subsubsection{Natural Generalized Mirage Mediation (nGMM)} 

The mirage mediation (MM) framework~\cite{mirage} is a well-motivated top-down
model that emerges from string theory where moduli fields are
stabilized via flux compactification. Within this framework, the soft
SUSY-breaking parameters typically receive comparable contributions from
moduli-mediated and anomaly-mediated contributions to SUSY breaking. As
a result, the superpartner spectrum is qualitatively different from that 
obtained in supergravity models. 
Since it was first proposed over a decade ago~\cite{kklt}, 
it has been shown that the scalar masses and $A$-parameters
are sensitive to the details of moduli stabilization and potential
uplifting mechanisms, while the gaugino mass pattern is robust~\cite{robust} 
with gaugino masses being given by
\be
M_a=\frac{m_{3/2}}{16\pi^2}\left(\alpha+b_ag_a^2\right).
\label{eq:mmg}
\ee
Here $m_{3/2}$ is the gravitino mass, and the phenomenological parameter 
$\alpha$ measures the relative strengths
of the moduli- and anomaly-mediated contributions to the gaugino masses.
The hallmark of MM models is that the measured values of gaugino
masses should apparently unify at the scale $$\mu_{\rm mir}= M_{\rm GUT}
\exp ^{\frac{-8\pi^2}{\alpha}},$$ while the gauge couplings unify, as
usual, at $M_{\rm GUT}$. There is no physical threshold at the scale
$\mu_{\rm mir}$, which has led to this mixed-modulus anomaly-mediation
framework being dubbed ``mirage unification''.  If $\mu_{\rm mir}$ is close
to the weak scale, the gaugino mass spectrum will be very compressed,
giving rise to a well-motivated top-down framework where winos (and
binos) can be essentially as heavy as gluinos.\footnote{If the auxiliary
field whose vacuum expectation value breaks SUSY transforms as a {\bf
75} dimensional representation rather than a singlet of $SU(5)$, the
weak scale gaugino masses have the ratio $M_1:M_2:M_3=6,6,-5$, also
resulting in a very compressed gaugino spectrum.}

It has been shown that the original MM models where scalar masses are 
determined from specific values of modular weights are very fine-tuned over
the entire range of parameters consistent with LHC sparticle and Higgs
mass constraints~\cite{siege}. 
In more general constructions, it was found that 
scalar masses and SUSY-breaking trilinear soft parameters are sensitive
to the details of the moduli stabilization and potential uplifting
mechanisms, while gaugino masses remain as in Eq. \ref{eq:mmg}. 
These considerations led us to propose a phenomenological generalization 
of the MM framework, hereafter referred to as Generalized Mirage Mediation (GMM), 
where scalar mass and trilinear soft terms are determined by 
continuous rather than discrete parameters, 
but where gaugino masses are still given by Eq.~(\ref{eq:mmg}). 
We refer the interested reader to Ref.~\cite{ngmm} for detailed
expressions for the GMM soft term values.
The parameter freedom of the GMM model allows for 
superpartner mass values yielding $\Delta_{\rm EW}$ as low as $\sim 15$ 
with $m_h\simeq 125$ GeV, but with much heavier bino and wino masses
(for a given value of $m_{\tg}$) as compared to models 
(such as NUHM2) with gaugino mass unification. 
Thus, in natural GMM, the SSdB signal as well as
the monojet plus soft dilepton signal (discussed in Sec.~\ref{nuhm2}) 
can become potentially unobservable even at the HL-LHC.
In such a case, natural SUSY may well elude the scrutiny of
HL-LHC; discovery of SUSY will then require new facilities.
The GMM model has been incorporated into the 
{\sc Isajet/Isasugra} computer code~\cite{isajet} which we use here 
to evaluate the superpartner spectrum as well
as for calculations of electroweak fine-tuning.

\subsubsection{The Natural Mini-Landscape}
\label{subsubsec:miniland}

The mini-landscape program is an attempt to target special regions of
the landscape of heterotic string theory 
(which allows for localized grand unification) 
where the effective low energy theory is the MSSM~\cite{miniland}. 
Detailed exploration in a series of papers has
led to a picture of the emergent 4-D theory whose properties are
determined by the geometry of the compact manifold and the location of
the matter and Higgs (super)fields on this manifold. The gauge group is
$SU(3)_{\rm col}\times SU(2)_L\times U(1)_Y$, though this symmetry may
be enhanced for fields confined to fixed points or fixed tori in the
compactified dimensions. Examination of models has led to the following
general picture~\cite{genpic}.
\begin{enumerate}
\item The first two generations of matter live on orbifold fixed points 
and exhibit a larger $SO(10)$ gauge symmetry with first and second generation
(s)fermions filling out the 16-dimensional spinor representation of
$SO(10)$.  
\item The Higgs multiplets, $H_u$ and $H_d$, live in the
untwisted sector and are bulk fields that feel just $G_{\rm SM}$.  These
fields, as well as the gauge multiplets, come in incomplete GUT
multiplets, automatically resolving the doublet-triplet splitting
problem.
\item The third generation quark doublet and the top singlet also
reside in the bulk and thus have large overlap with the Higgs fields
and correspondingly large Yukawa couplings. The small overlap of Higgs
and first and second generation fields (which do not extend into the bulk)
accounts for their much smaller Yukawa couplings. The location of other
third generation matter fields is model-dependent. For simplicity, we
assume that they also to live in the bulk.
\item Supergravity breaking may arise from hidden sector gaugino
condensation with $m_{3/2}\sim \Lambda^3/m_{\rm Pl}^2$, with the gaugino
condensation scale $\Lambda\sim 10^{13}$ GeV. SUSY breaking effects are
felt differently by the various MSSM fields, depending on their
location. The Higgs and stop fields in the untwisted sector feel
extended supersymmetry (at tree level) in 4-dimensions, and are thus
more protected than the fields on orbifold fixed points which receive
protection from just $N=1$ supersymmetry~\cite{Krippendorf:2012ir}.
First and second generation matter scalars are thus expected with masses
$\sim m_{3/2}\sim 10-30$ TeV. 
Third generation and Higgs soft mass parameters (which
enjoy the added protection from extended SUSY) are suppressed by an
additional loop factor $\sim 4\pi^2 \sim \log (m_{\rm Pl}/m_{3/2})$
typically expected in MM models.  Gaugino masses and third generation
trilinear soft terms are suppressed by the same factor, leading to 
TeV-scale mass values as expected from the nGMM framework discussed above. 
The main difference is that while all matter field generations were 
assumed comparable in the GMM framework, 
the mini-landscape picture distinguishes the third
generation from the much heavier first two generations.
\end{enumerate}

Several aspects of the spectrum of these models are qualitatively
similar to
that of the GMM models. The main difference is that while the three
generations of matter scalars are treated identically in the GMM
framework (except, of course, for the difference in Yukawa interactions);
in the mini-landscape picture, the first and second generation sfermions are
considerably heavier than their third generation cousins.  In this
sense, the natural mini-landscape picture is closer to the NUHM3 RNS
model, but differs from it because the gaugino mass pattern is
determined by MM rather than by gaugino mass unification.  The broad brush
phenomenology of the natural mini-landscape picture has been laid out in
Ref.~\cite{minilandpheno}, to which we refer the reader for details. We
have used Isajet/Isasugra~\cite{isajet} for the evaluation of the superpartner
spectrum, as well as for $\Delta_{\rm EW}$, in this framework. 
For our purposes, the gaugino spectrum is compressed for viable regions of model
parameter space, so that, as in the nGMM model, the SSdB and monojet plus
soft dilepton signals may be inaccessible, and we may have to 
rely on gluino and stop signals for definitive detection.

\subsection{The Reach of LHC33 for Natural SUSY Models}
\label{subsec:glstop}

We have already mentioned that in the natural NUHM2 model, the SSdB
signal from wino pair production, together with the monojet plus soft
dilepton signal, are sufficient to probe essentially the entire
$\Delta_{\rm EW}< 30$ region of the phenomenologically viable parameter
space of the model at the HL-LHC. As already stressed, the model
characteristics crucial for arriving at this conclusion are 1)~gaugino
mass unification and 2)~the naturalness upper limit $m_{\tg} \alt
5$~TeV on the gluino mass,
which implies a corresponding limit on the wino mass. The existence of
well-motivated natural SUSY models where one of these conditions is not
satisfied was one of our main motivations for studying the gluino and top
squark reach at LHC33.

\begin{figure}[tbp]
\begin{center}
\includegraphics[width= 16 cm,clip]{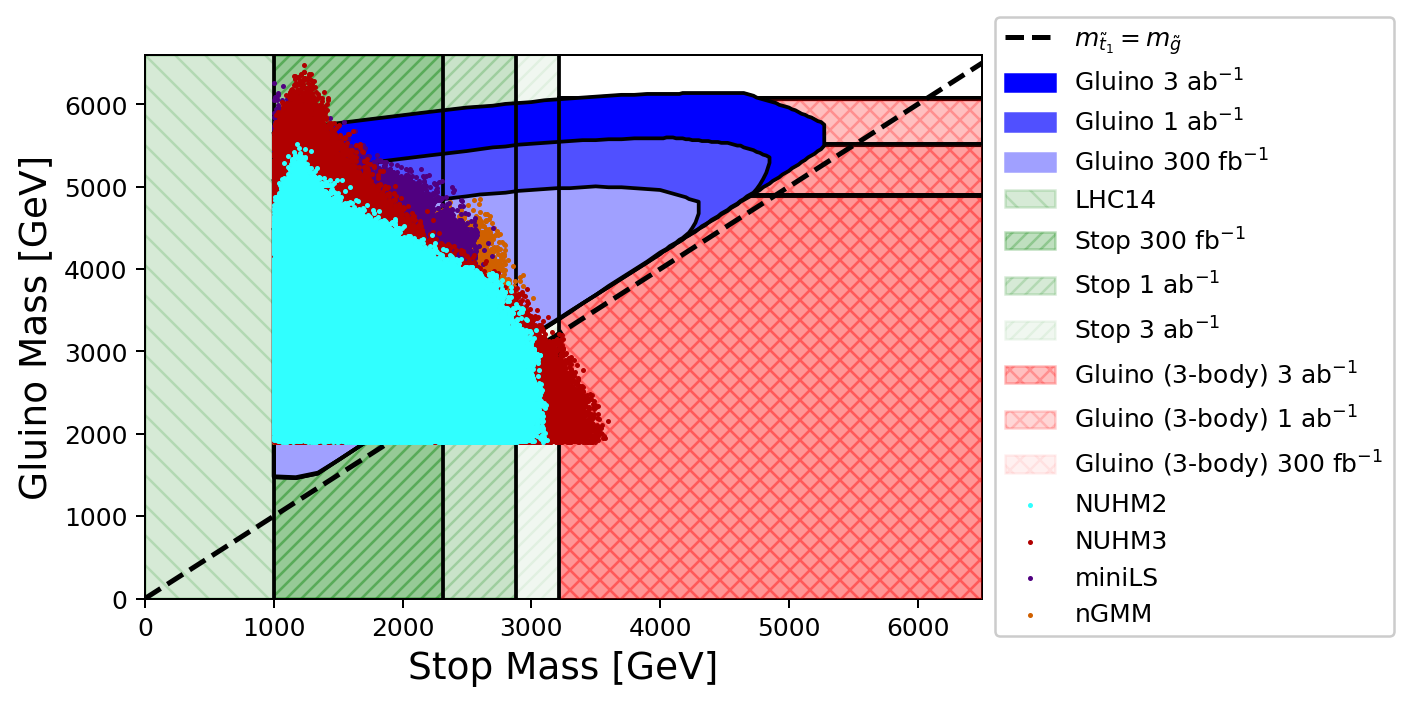}
\caption{ Points with $\Delta_{\rm EW}<30$ in the stop-gluino mass plane from 
dedicated scans of the NUHM2, NUHM3, nGMM and mini-landscape models 
described in Sec. \ref{sec:implications}.
These are plotted along with (i)
the projected $5\sigma$ stop  reach of LHC33 from Fig.~\ref{fig:stop-reach},
(ii) the projected $5\sigma$ gluino reach of LHC33 from
Fig.~\ref{fig:gluino-reach}  obtained
assuming that the gluino decays via $\tg \to
\tst_1 \bar{t}$, $\tst_1^\ast t$, and (iii) the projected $5\sigma$ LHC33 reach
for gluino pairs from Fig.~\ref{fig:3-body-simplified-reach} obtained
assuming that the gluino decays via off-shell
stops and sbottoms. 
The 95\% CL exclusion lines are typically $400-500$ GeV higher. 
\label{fig:scatter}}
\end{center}
\end{figure}

With this in mind, we begin by showing a scatter plot of the stop versus
gluino mass in Fig.~\ref{fig:scatter} for the four models
discussed in Sec.~\ref{subsec:rns}: natural NUHM2 (cyan/ lightest), 
natural NUHM3 (dark red/ second darkest), 
natural mini-landscape (purple/ darkest)
and nGMM (orange/ second lightest).  
For each model, we have scanned
the parameter space consistent with the Higgs mass as well as LHC
sparticle mass constraints ($m_{\tg}>1.9$ TeV) 
and placed a symbol in the figure if we find
a solution compatible with $\Delta_{\rm EW} < 30$. For the sake of
brevity, we will not describe the details of the scans here, but refer
the reader to our earlier studies of the NUHM model~\cite{lhc}, the nGMM
model~\cite{ngmm}, and the natural mini-landscape picture
~\cite{minilandpheno}, where the parameter space of each model is also
detailed.\footnote{Aside from the broad scans that yield a wide
range of $\Delta_{\rm EW}$ in these papers, we have
  also carried out focussed scans targetting the low $\Delta_{\rm EW}$
  region to obtain accurate naturalness bounds on stop and gluino masses.}
Also shown are the regions of the stop-gluino mass plane accessible via
searches for stop or gluino pair production at LHC33 using the
strategies described in Sec.~\ref{sec:reach}.\footnote{For the case
  $m_{\tst_1}+m_t > m_{\tg}$, we assume that the reach is essentially
  independent of $m_{\tst_1}$. We expect this to be the case except,
  perhaps, very close to the kinematic boundary for gluino two-body decays.}

Since both gluinos and stops decay essentially with the branching
fractions assumed in our simplified model analyses of
Sec.~\ref{sec:reach}, the reach results that we obtained there should be
directly applicable.  In particular, LHC33 experiments should be able to
discover stops with a $\ge 5\sigma$ significance for $m_{\tst_1}<
2.3$ (2.8) [3.2]~TeV, assuming an integrated luminosity of 0.3 (1.0)
[3.0]~ab$^{-1}$: {\it i.e.}, to the left of the correspondingly labelled
lines in the figure.  Also shown are the corresponding
discovery contours for the gluino. For heavy stops, these extend to
5.5~TeV for an integrated luminosity of 1~ab$^{-1}$ and beyond 5~TeV if
the stop is just 1~TeV (in which case it would be easily discovered 
at the $14$ TeV LHC). We have not shown the 95\% CL contours here,
but these exclusion limits typically extend $400-500$~GeV beyond the
discovery reaches shown in the figure.

The following features of Fig.~\ref{fig:scatter} are particularly worth
noting:

\begin{enumerate}

\item We see that at least one of the gluino or the stop should be
  discoverable at LHC33 even with an integrated luminosity of
  300~fb$^{-1}$ for all viable models with $\Delta_{\rm EW} < 30$. 

\item For the anticipated LHC33 integrated luminosity of 1~ab$^{-1}$,
  there should be a signal in {\em both} gluino as well as stop pair
  production channels for most of the parameter space with $\Delta_{\rm
  EW}< 30$. 

\item If an integrated luminosity of 3~ab$^{-1}$ (the integrated luminosity
  expected at the HL-LHC) is accumulated at LHC33, both gluino and stop
  signals should be observable for essentially all natural SUSY models,
  the exceptions being the regions with either $m_{\tst_1}< 1.3$~TeV (in
  which case the stop should have been already discovered at the
  HL-LHC), or the region with $m_{\tg} < 3.2$~TeV, much of which would
  be probed at the HL-LHC where the 5$\sigma$ reach extends to 2.8~TeV.

\item As mentioned above, $m_{\tg}$ is bounded above for all
models. This bound is roughly model-independent for values of
$m_{\tst_1} \agt 2$~TeV, but for lighter stops, $m_{\tg}$ can be
significantly larger, extending to as high as 6~TeV especially in the
NUHM3 and the mini-landscape models where the first and second
generation sfermions are much heavier than the third generation
sfermions. For these models, the increase in stop masses from heavy
gluino loops (which results in too large of a $\Delta_{\rm EW}$) is
compensated by negative two loop contributions due to heavy
first/generation. This effect is clearly smaller in models where the
high scale third generation mass is not independent of other scalar
masses, resulting in a tighter naturalness bound on $m_{\tg}$.

\end{enumerate}

The take-away message from this figure is that over the entire range of
natural SUSY models with $\Delta_{\rm EW} < 30$, experiments at LHC33,
with an integrated luminosity of 1~ab$^{-1}$, will have the sensitivity
to discover at least one of the stop or the gluino -- 
and over much of the parameter space, both. In models with gaugino mass
unification, there will likely be observable signatures in the SSdB
channel\footnote{In Ref.~\cite{lhc} we had projected that the HL-LHC
would probe the entire natural parameter space via the SSdB
signature. Since then, we have discovered small additional backgrounds
and also (via more dedicated scans) found that the gluino mass bound
extends somewhat beyond $4.5$~TeV (see Fig.~\ref{fig:scatter}) that we
projected in our earlier study. HL-LHC experiments should nonetheless
be able to discover SUSY over much of the parameter space of this
model.} and perhaps also in other channels because the $\tz_2$ decays
will lead to an excess of soft dilepton pairs with $m_{\ell\ell} <
m_{\tz_2}-m_{\tz_1}$ in events triggered by hard jets or large $\eslt$.

The reader may be concerned that the analysis that we have presented is
specific to the models examined in the study. We should mention that
these models encompass a wide range of models with the superpartner
content of the MSSM.  Our results for the reach for gluinos and stops
are unlikely to be significantly altered in natural SUSY models from our
projections, since these projections depend essentially on the
production cross sections and on the decay patterns. The latter depend
on the existence of light higgsinos and on the fact that third
generation squarks are significantly lighter than other
squarks.\footnote{If first/second generation squarks are also light,
their production would enhance the SUSY signal and likely offer
additional channels to search for SUSY.} It is also difficult to
envision how the upper limits on gluino and stop masses obtained from
$\Delta_{\rm EW} < 30$ could be significantly altered from those in the
NUHM3 RNS model, where the high scale third generation squark and gluino
masses can be chosen independently. The NUHM3 bound on $m_{\tg}$ is
unlikely to be very sensitive to the gaugino mass unification
assumption. This is confirmed by the fact that the distributions of
points for the NUHM3 model and for the mini-landscape model are
qualitatively similar. We therefore conclude that LHC33 experiments
should be able to discover natural SUSY (as we have defined it) in all
models with a minimal superpartner spectrum.  The absence of any
signal\footnote{Recall that the 95\% CL exclusion contours are typically
0.5~TeV beyond the 5$\sigma$ discovery contours in Fig.~\ref{fig:scatter}.}
will mean that even though weak scale SUSY may resolve the big hierarchy
problem, a little hierarchy would remain ({\it i.e.} SUSY would contain
significant electroweak fine-tuning).

\section{Summary and Concluding Remarks}
\label{sec:conclusions}

We have examined the proposed LHC33 reach for SUSY via stop and gluino
pair production signals (assuming that higgsinos are the lightest SUSY
particles with masses not much larger than $m_Z$) assuming an integrated
luminosity of $\sim 0.3-3$~ab$^{-1}$.  For our analyses, we assume that
the stop decays to higgsino-like charginos and neutralinos via
$\tst_1\to t\tz_{1,2}$ and $\tst_1 \to b\tw_1$, and that the gluino
decays via $\tg\to t\tst_1$ or via the three-body decays $\tg\to
t\bar{t}\tz_{1,2},tb\tw_1$. In this study, we conservatively assume that
the decay products of the higgsino-like $\tw_1$ and $\tz_2$ are too soft
to be detected. 

We believe that our analysis with the assumed stop and gluino decay
patterns is strongly motivated by rather conservative naturalness
arguments that imply spectra with gluinos may be as heavy as 6~TeV and
light stops as heavy as 3~TeV can be natural as long as higgsinos are
lighter than $\sim 350$~GeV.  Gluino and stops may then be well beyond
the discovery reach of even the HL-LHC which extends to $m_{\tg} \alt
2.8$~TeV and $m_{\tst_1} < 1.2$~TeV, while signals from electroweak
production of higgsinos may be buried below SM backgrounds if these are
essentially degenerate. Indeed, there are  theoretically
well-motivated natural SUSY models with compressed gaugino spectra (see
Sec.~\ref{subsec:rns}) where SUSY may completely elude detection at the
HL-LHC. Readers who do not subscribe to these naturalness arguments
should view our study as an analysis of the SUSY reach of the proposed
energy upgrade of the LHC for models with light higgsinos.

Our optimized cuts to extract the stop production signal at LHC33 are
listed in (\ref{stop_cuts_final}) while the corresponding cuts for
the gluino production signal are shown in (\ref{gluino_cuts_final}).
From our results for the stop cross section after cuts shown in
Fig.~\ref{fig:stop-reach}, we project that LHC33 experiments should be
able to discover stops at $5\sigma$ level with a mass values 
up to 2.3 (2.8) [3.2]~TeV, assuming 
an integrated luminosity of 0.3 (1) [3]~ab$^{-1}$. If no signal is seen,
the 95\% CL exclusion limits extend about 600~GeV further. The gluino
reach of LHC33 is illustrated in Fig.~\ref{fig:gluino-reach}  assuming
that gluinos decay via $\tg\to t\tst_1$  and in  
Fig.~\ref{fig:3-body-simplified-reach} for the case that gluinos decay
to third generation quarks plus higgsinos. In both cases we find that
LHC33 experiments should be able to discover a gluino 
at $5\sigma$ level for $m_{\tg}$ as heavy as 
5 (5.5) [6]~TeV assuming an integrated luminosity of 0.3 (1) [3]~ab$^{-1}$. 
The 95\% CL exclusion reach values are typically 400-500 GeV higher 
than the $5\sigma$ values.

In Sec.~\ref{sec:implications}, we have examined the implications of our
gluino and stop reach results for natural SUSY models, defined to have
no worse than 3\% electroweak fine-tuning. We have stressed that
electroweak naturalness is a very conservative approach to fine-tuning
in that it allows for the possibility that underlying model parameters
(usually taken to be independent) might be correlated in the underlying
theory. Disregarding this may prematurely cause us to discard perfectly
viable models because they appear to be unnatural in an effective
theory. Even with this conservative approach, we find that experiments
at LHC33 will be able to discover either the stop or the gluino even
with an integrated luminosity of 300~fb$^{-1}$ in {\em all natural SUSY
models} where the low energy theory is the MSSM. With an integrated
luminosity of 1-3~ab$^{-1}$, for the bulk of the models, both gluino and
stop signals should be observable, together perhaps with some additional
signals for specific scenarios: see Fig.~\ref{fig:scatter}. This is in
sharp contrast to the HL-LHC, where signals from natural SUSY models
with compressed gaugino masses may well remain undiscovered because
gluinos, stops, and winos are too heavy and higgsino signals are buried
under SM backgrounds.  LHC33 will, therefore, allow an unambiguous test
of natural SUSY models, conservatively defined by spectra with no worse
than 3\% electroweak fine-tuning.

\section*{Acknowledgments}

We acknowledge useful communications with T. Cohen. 
This work was supported
in part by the US Department of Energy, Office of High Energy Physics;
was aided by the use of SLAC Computing Resources;
and was performed in part at the Aspen Center for Physics, 
which is supported by National Science Foundation grant PHY-1607611.
%

\end{document}